\documentclass{jaa}

\usepackage{graphicx}
\usepackage{caption}
\usepackage{subcaption}

\begin{document}\sloppy

\title{A Search for variable stars in the four open
star clusters}

\author{Alok Durgapal\textsuperscript{1}, Geeta Rangwal\textsuperscript{1,*}, D. Bisht\textsuperscript{2}, Harmeen Kaur\textsuperscript{1}, R. K. S. Yadav\textsuperscript{3} and J. C. Pandey\textsuperscript{3}}
\affilOne{\textsuperscript{1}Department of Physics, DSB Campus, Kumaun University, Nainital, India.\\}
\affilTwo{\textsuperscript{2}Key Laboratory for Researches in Galaxies and Cosmology, University of Science and Technology
             of China, Chinese Academy of Sciences, Hefei, Anhui 230026, China.\\}
\affilThree{\textsuperscript{3}Aryabhatta Research Institute of Observational Sciences, Manora Peak, Nainital, India.}


\twocolumn[{

\maketitle

\corres{geetarangwal91@gmail.com}

\msinfo{......}{........}

\begin{abstract}
We present a CCD photometric survey for the search of variable stars in 
four open clusters namely  Berkeley 69, King 5, 
King 7,  and Berkeley 20.  
The time series observations were carried out for 1 and/or 2 nights 
for each of the clusters in the year 1998, which have led to 
identify nineteen variable stars in these clusters. Out of these 
19 variable stars, five stars show  $\delta$ Scuti like variability 
and two stars show W UMa type variability. In other stars, we could not 
find the periods and hence the type of variability due to the lack of 
sufficient data. The periods of $\delta$ Scuti type stars are found to 
be in the range of 0.13 to 0.21 days, whereas the two stars in the cluster 
Berkeley 20, which showed W UMa type variability have orbital periods of 0.396 
and 0.418 days, respectively.  Using the Gaia data, the basic parameters 
of the clusters Berkeley 69, King 7 and King 5 are also revised. 
The age and reddening 
are estimated to be $0.79\pm0.09$ Gyr and $0.68\pm0.03$ mag for Berkeley 69, $0.79\pm0.09$ 
Gyr and $1.22\pm0.03$ mag for the cluster King 7 and $1.59\pm0.19$ Gyr 
and $0.63\pm0.02$ mag for the cluster King 5, respectively. Signature
of mass segregation is found in the clusters King 7 and King 5. 

\end{abstract}


}]



\doinum{12.3456/s78910-011-012-3}
\artcitid{\#\#\#\#}
\volnum{000}
\year{0000}
\pgrange{1--}
\setcounter{page}{1}
\lp{1}

\begin{figure*}
\begin{subfigure}{1.0\columnwidth}
\centering
{\includegraphics[width=\columnwidth,trim={0.0cm 0.0cm 0.7cm 3.8cm}]{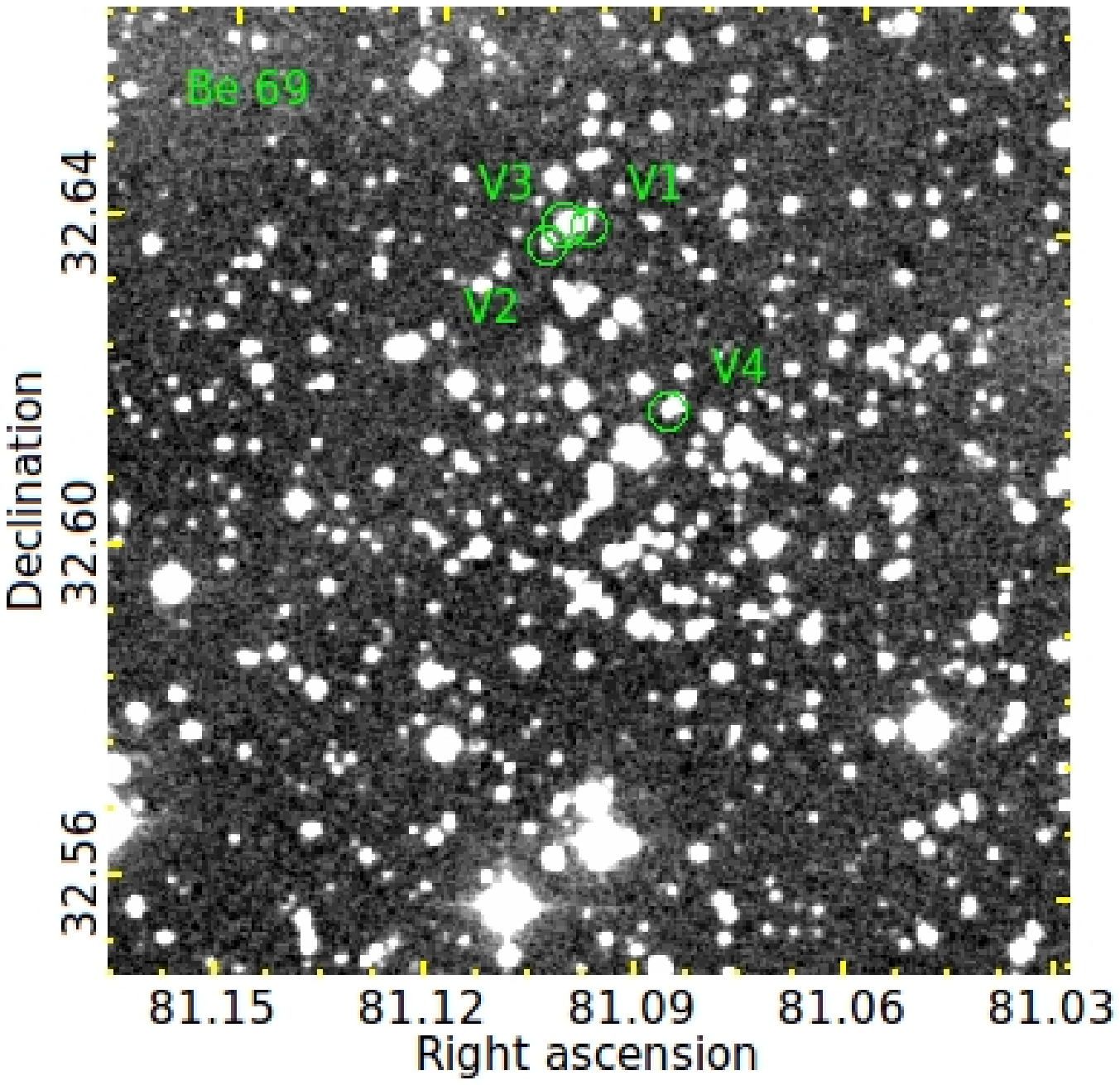}}
\caption{Berkeley 69}
\label{id_69}
\end{subfigure}
\begin{subfigure}{1.0\columnwidth}
\centering
{\includegraphics[width=\columnwidth,trim={0.0cm 0.0cm 0.7cm 3.8cm}]{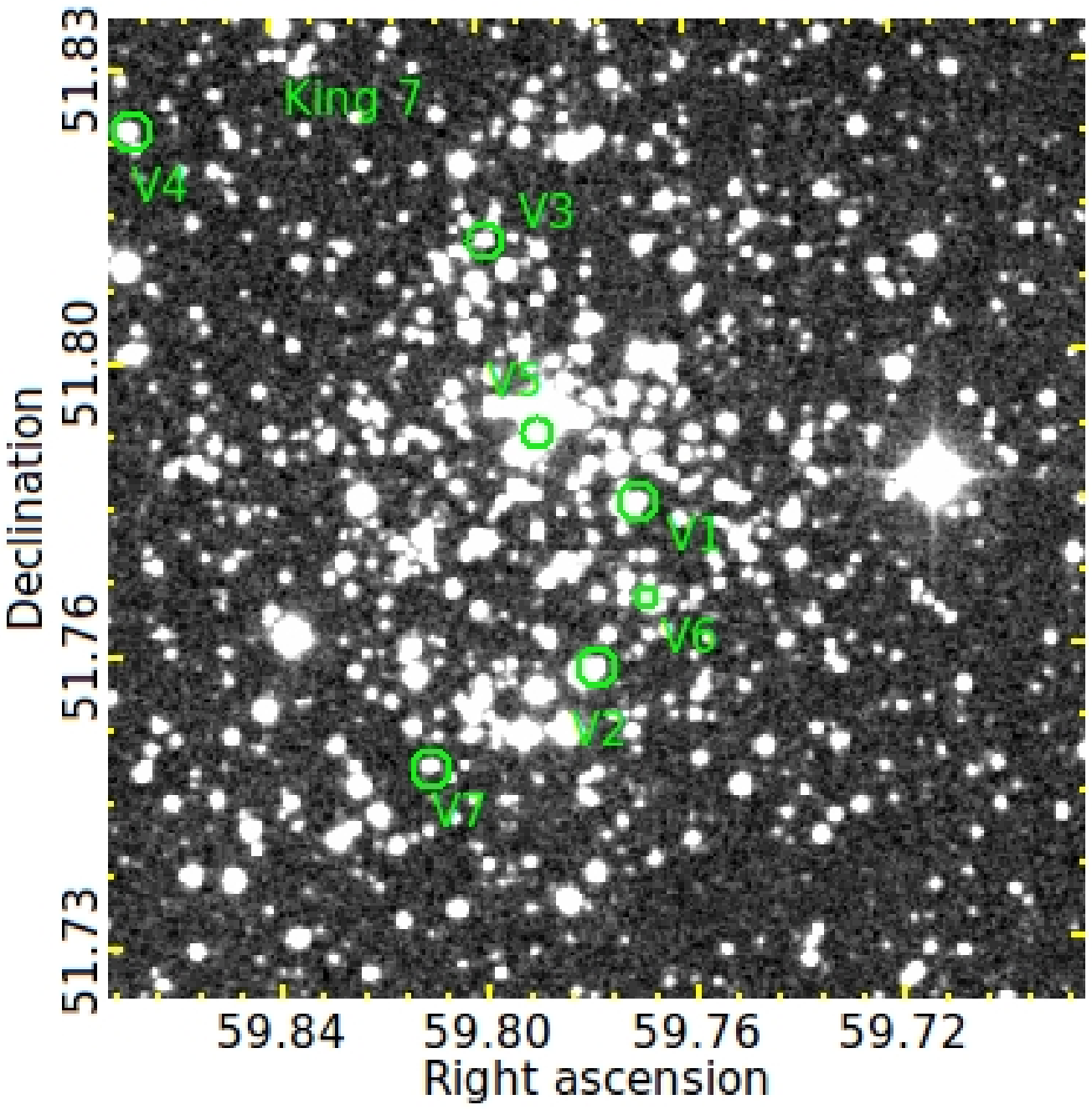}}
\caption{King 7}
\label{id_7}
\end{subfigure}
\begin{subfigure}{1.0\columnwidth}
\centering
{\includegraphics[width=\columnwidth,trim={0.0cm 0.0cm 0.7cm 0.3cm}]{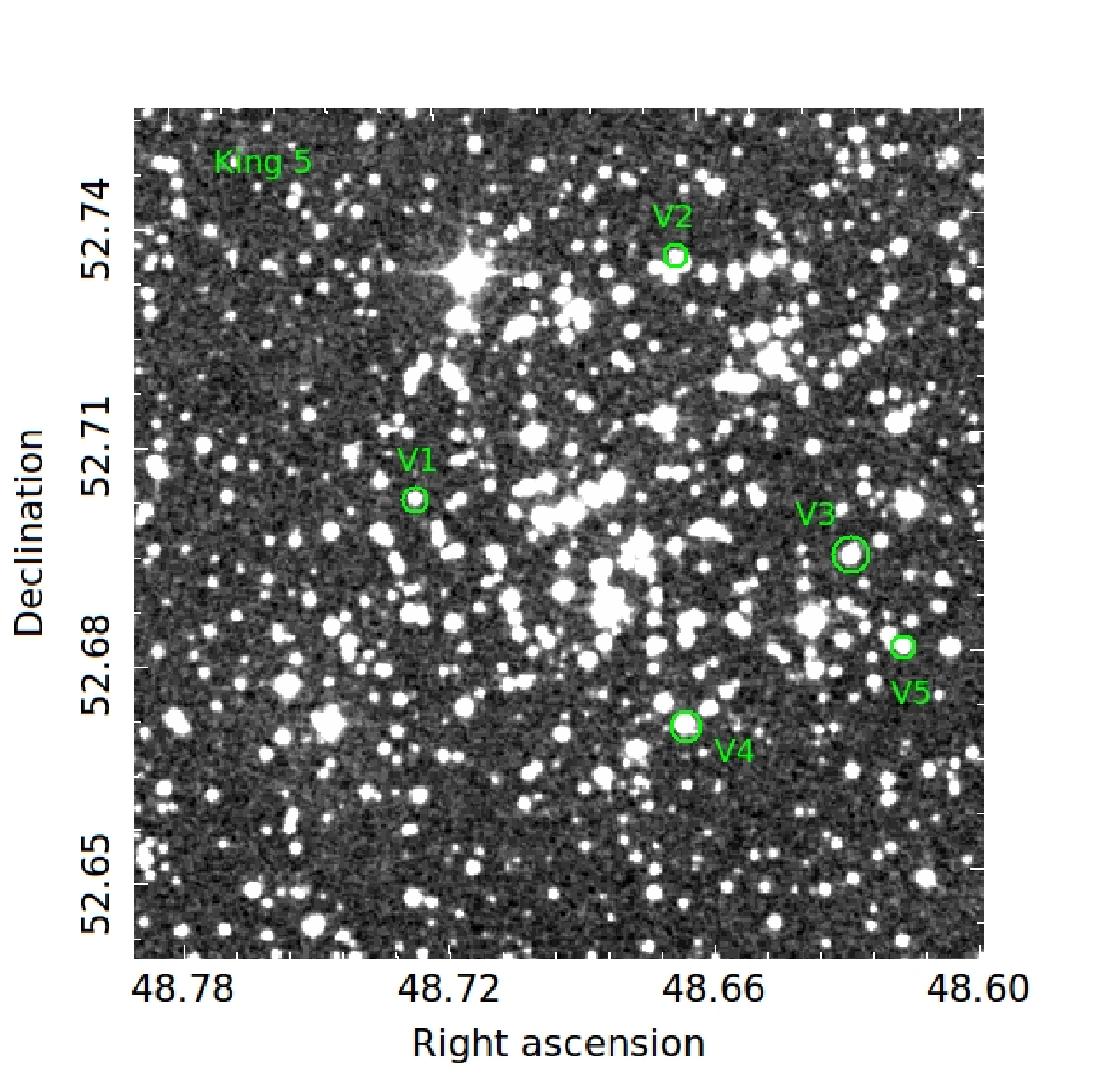}}
\caption{King 5}
\label{id_5}
\end{subfigure}
\begin{subfigure}{1.0\columnwidth}
\centering
{\includegraphics[width=\columnwidth,trim={-1.0cm 0cm 0.7cm 0.3cm}]{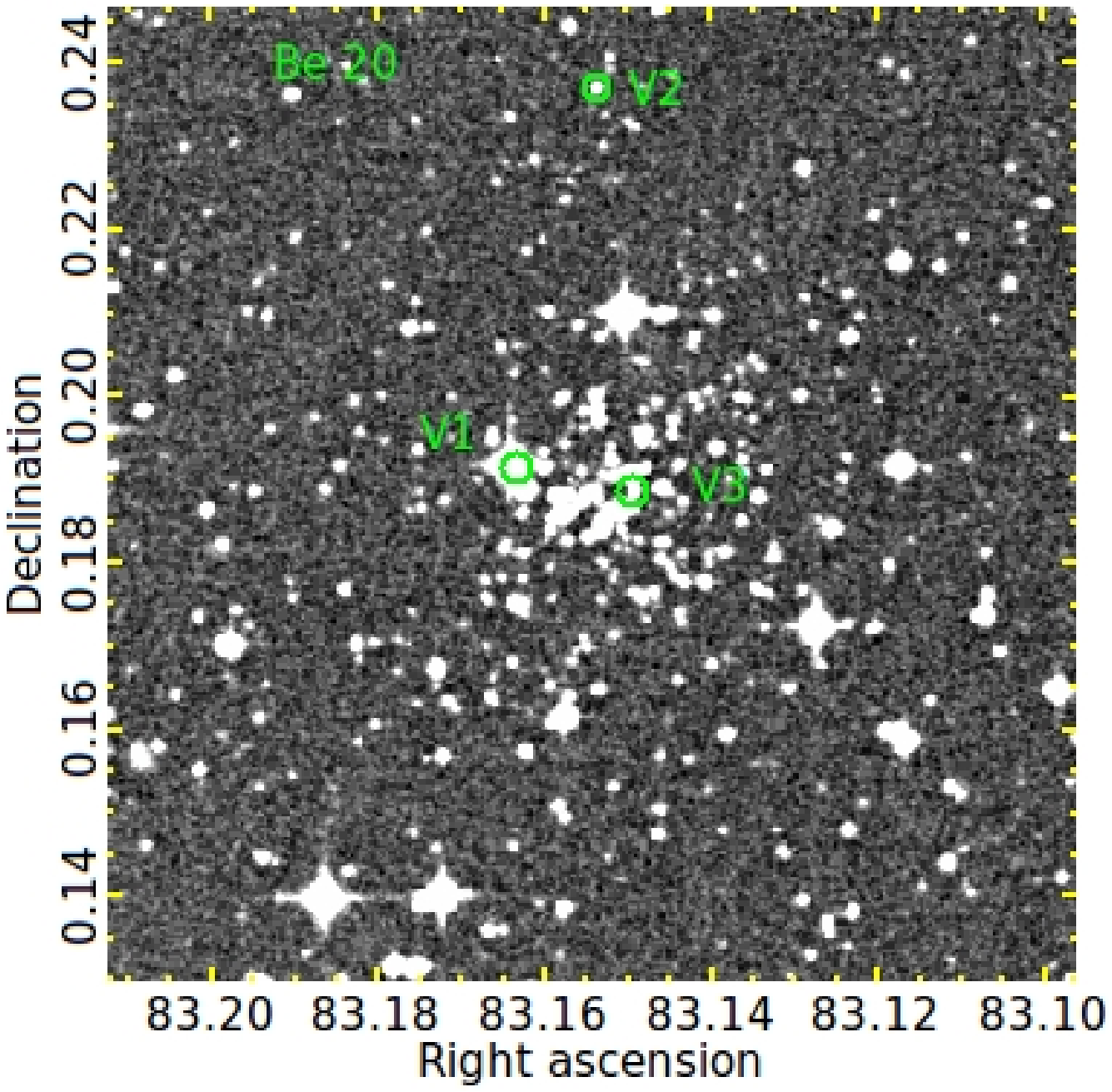}}
\caption{Berkeley 20}
\label{id_20}
\end{subfigure}
\caption{ The R band $7^{\prime} \times 7^{\prime}$ image of the clusters Berkeley 69, King 7, King 5 and Berkeley 20 taken from DSS.  Variable stars identified in the cluster region are encircled.}
\label{id}
\end{figure*}

\section{Introduction}\label{int}
Variable stars play an important role to understand the stellar structure and evolution.
The variable stars in open clusters are more useful because their parameters e.g., reddening, age, and distance are better known in comparison to those in the field region.
It would be very important for observational studies of
multi periodic pulsating variable stars such as $\delta$ Scuti stars, $\gamma$ Dor
stars and SPBS (slowly pulsating B stars) etc, because they require
accurate time series data to analyze their complicated light curves.
There is currently much interest shown by various groups to search
the variable stars in open clusters (see e.g. Jeon {\em et al.} 2016;
Popov {\em et al.} 2017; Hutchens {\em et al.} 2017; Schaefer {\em et al.}
2018; Smith {\em et al.} 2019; Yepez {\em et al.} 2019).
Further, One of the most effective and productive usages of the 1-m class
telescope may be CCD time series photometry of variable stars in star
clusters (see e.g. Joshi et al. 2012, Lata et al. 2014, 2016, 
Dutta et al. 2019). Simultaneous CCD photometry of all stars in the cluster
enables us to do the efficient observations. Moreover, one can obtain
precise time series data by observing the stars under the same weather
conditions. 

\begin{figure*}
\includegraphics[width=\columnwidth]{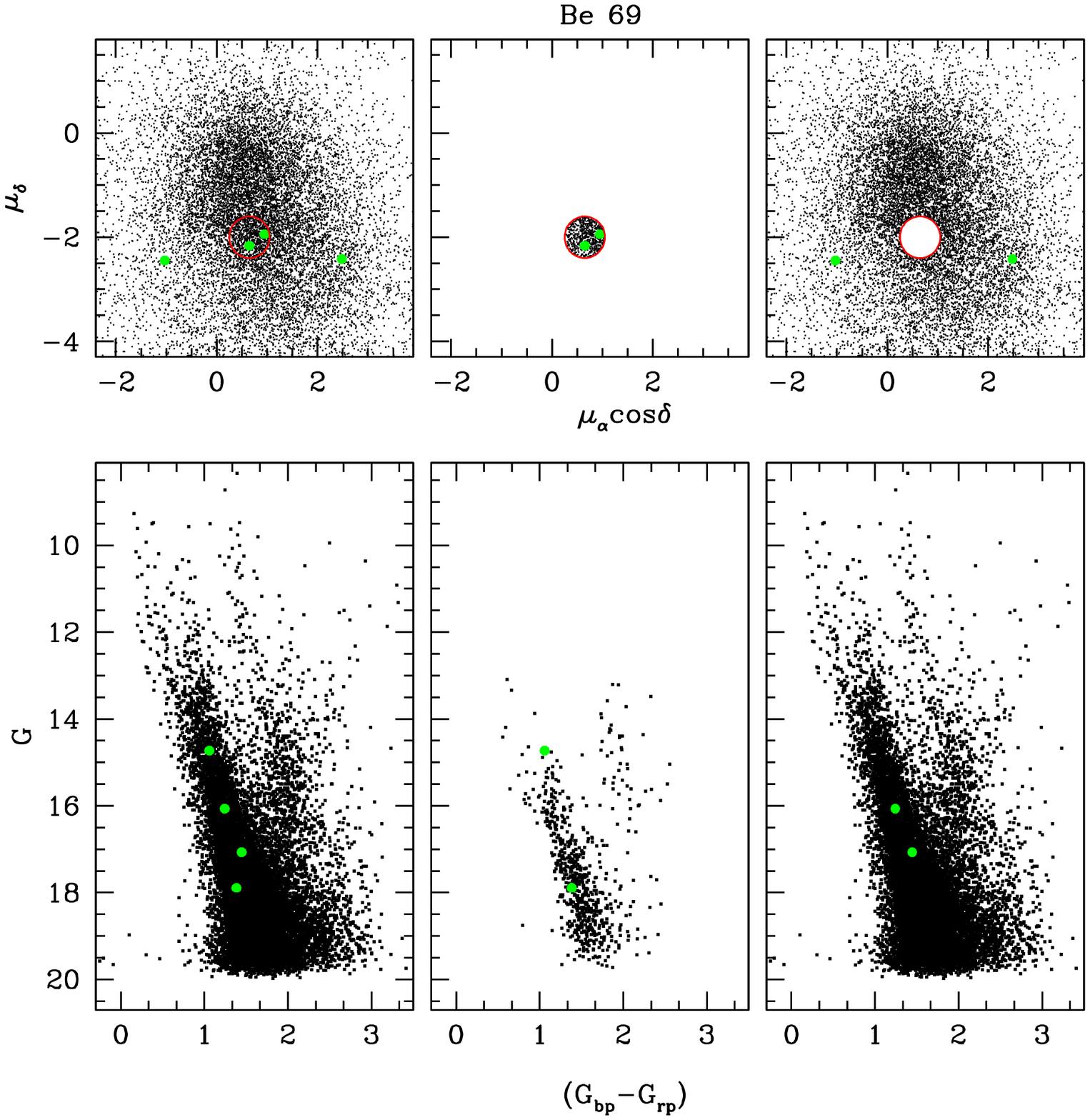}
\includegraphics[width=\columnwidth]{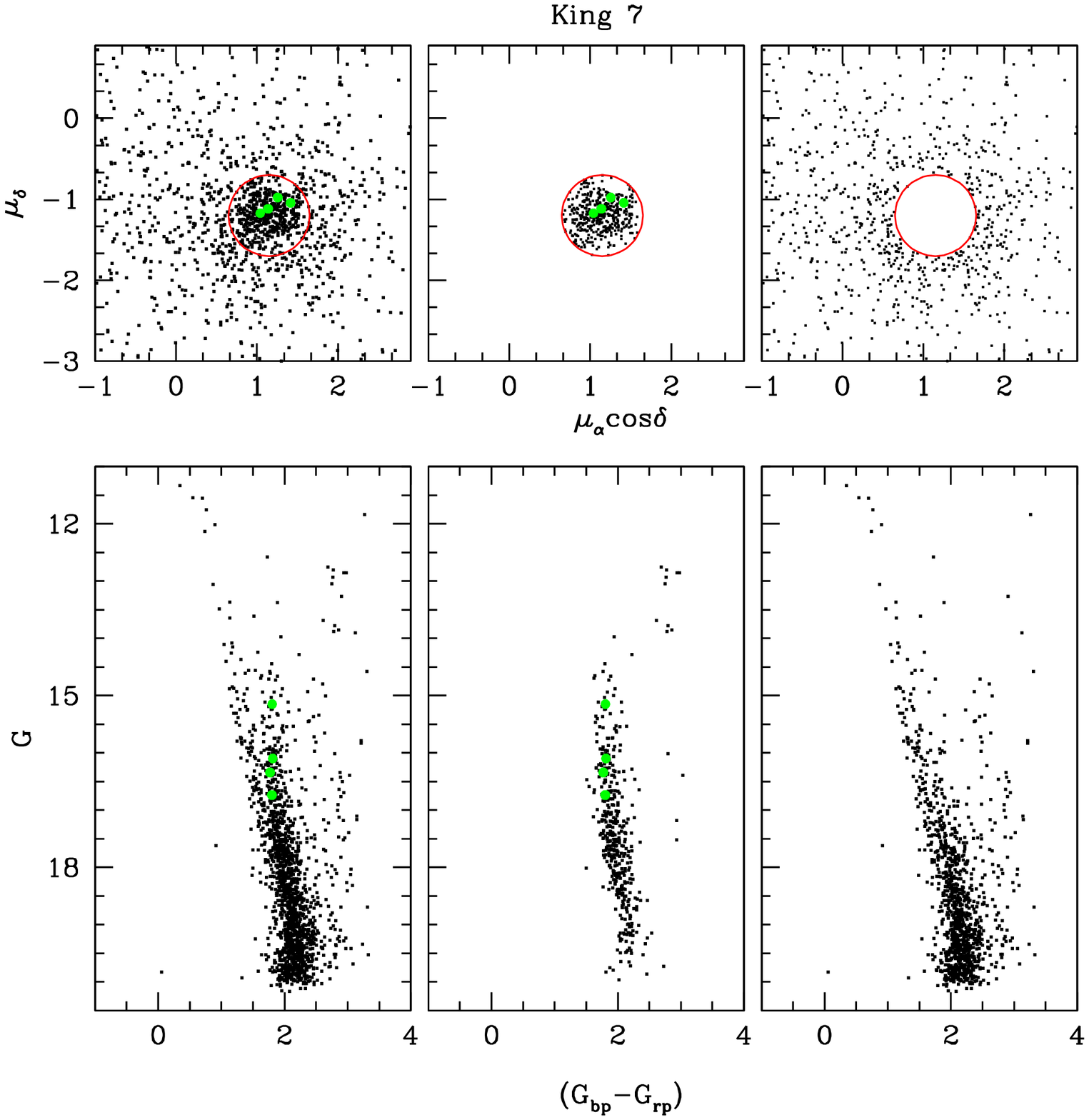}
\caption{Proper motion vector-point diagram (upper panels) and $(G_{bp}-G_{rp}), G$
CMD (lower panels) for the clusters Berkeley 69 and King 7. Left panels show the complete
sample of the cluster, middle panels shows most probable cluster members
and right panels show the field stars. Green points represent the variable
stars identified in our analysis.}
\label{vpd_7}
\end{figure*}

We have therefore searched for the variable stars in four northern intermediate age clusters King 5, King 7, Berkeley 20, and Berkeley 69. Previously, we have studied these clusters  in detail using
CCD U, B, V, R, and I photometric data taken from 104-cm Sampurnanand telescope,
of ARIES, Nainital (Durgapal {\em et al.} 1996; Durgapal {\em et al.} 1997;
Pandey {\em et al.} 1997; Durgapal {\em et al.} 1998; Durgapal \& Pandey
2001; Durgapal 2001; Durgapal {\em et al.} 2001). In our previous papers, we have determined the fundamental
properties of these clusters and studied their dynamical state. Apart from our 
analysis other studies for these clusters have also been carried out 
in the past. Such as
Bukowiecki \& Maciejewski (2008) searched variable stars in the region of 
King 7 and found a total of 16 variable stars. Out of which only five
stars lie within the cluster's tidal radius.
Oralhan {\em et al.} (2015) studied King 5, using CCD $UBV(RI)_{C}$
photometric data from the 84-cm telescope and determined the fundamental
astrophysical parameters, $E(B-V)$ as 0.70 mag, Z=0.01
and heliocentric distance as 1.74 kpc.
Hoq \& Clemens (2015) have studied
King 5 and King 7 using 2MASS data in near-IR wave-band and derived the
fundamental parameters as log(age)=$9.15 \pm 0.12$, distance = 1.87 kpc, 
$E(B-V)= 0.71 \pm 0.08$ mag 
for king 5 and log(age)=$8.03 \pm 0.12$, distance = 3.82 kpc, 
$E(B-V)=1.46 \pm 0.07$ for the cluster King 7. These results 
are in good agreement
with the results reported by us earlier. The present work is an effort to
detect the variable stars in the above mentioned open clusters. We have also revisited the
fundamental properties of King 5 and King 7 using the Gaia DR2 data.
General information about the clusters under study are listed in the 
Table \ref{fp}, which are taken from the WEBDA database.

Structure of the article is as follows. In section \ref{obs} we discussed
the observational data and its reduction. Member selection method and criteria
are discussed in \ref{kin} In section \ref{fpara} we determined the
fundamental parameters of the clusters using Gaia data. We discussed
the variable stars detected in the clusters in the section \ref{var} Mass
segregation in the clusters is discussed in section \ref{mass}
At last we concluded our analysis in section \ref{con}

\begin{table*}
   \centering
   \caption{Fundamental parameters of Berkeley 69, King 7,
    King 5 and Berkeley 20 taken from WEBDA database, where $d$ is the
    heliocentric distance of the clusters.}
   \begin{tabular}{ccccccccc}
   \hline\hline
  Cluster & $\alpha$ & $\delta$ & $l$ & $b$ & $d$ & $log(t)$ & $E(B-V)$ & [Fe/H] \\
          & (J2000) & (J2000) & degree & degree & kpc &   & mag &    \\
  \hline
   Berkeley 69  & 05:24:36 & +32:39:00 & 174.435 & -01.787 & 2.86 & 8.95 & 0.65 & - \\
   King 7 & 03:59:00 & +51:48:00 & 149.774 & -01.019 & 2.20 & 8.80 & 1.25 & - \\
   King 5 & 03:14:45 & +52:41:12 & 143.776 & -04.287 & 1.90 & 9.00 & 0.76 & -0.30  \\
   Berkeley 20 & 05:33:00 & +00:13:00 & 203.483 & -17.373 & 8.40 & 9.78 & 0.12 &-0.61  \\
  \hline
  \end{tabular}
  \label{fp}
  \end{table*}

\section{Observations and  data reduction}\label{obs}

We have observed four clusters (Berkeley 69, King 7, King 5 and Berkeley 20) in
$V$ band for about 5-6 hours on six nights during October - December 
1998 using a CCD camera mounted on 104-cm Sampurnanand reflector telescope 
of ARIES, Nainital. The CCD camera consists of  $1024\times 1024$ pixels 
with each pixel size of  24 $\mu$. The field of view of the CCD is 
$\sim$6.0  $\times 6.0$ arcmin$^2$, where each square pixel covers a 
$0.37\times 0.37$ arcseconds$^2$ of the sky. The  read-out-noise and gain  
of the CCD are 7.0 e$^-$ and  11.98 e$^-$per ADU, respectively.  
In order to improve the S/N ratio, the observations were taken in the 
binning mode of $2 \times 2$ pixels. All the observations have been  taken 
in the V filter with an exposure time of 150 sec. Log of the observations 
are listed in Table \ref{expo}.
Bias and flat-field frames were also taken along with the science frames 
for pre-reduction processes. The standard process of image cleaning was 
employed using ESO-MIDAS software. The photometry of cleaned cluster images 
was done by using DAOPHOT package by Stetson (1987). Since we are
using the differential magnitudes of the stars in the present analysis,
we performed only aperture photometry. 
The identification charts for the clusters Berkeley 69, King 7, King 5, and Berkeley 20 
are shown in Figure \ref{id} and the variable stars identified by us are also 
shown by open circles. 

\begin{table*}
   \centering
   \caption{Details of the observations for the clusters Berkeley 69, King 7,
    King 5 and Berkeley 20} 
   \begin{tabular}{ccccccccc}
   \hline\hline
  Cluster & Filter & Exposure time & Date & Observing span \\
          &        & (sec)         &      & (sec)    \\
  \hline
   Berkeley 69  & V & $150 \times 086 $ & 09 Dec 1998 & 12900   \\
                &   & $150 \times 042 $ & 25 Dec 1998 & 06300   \\
   King 7 & V & $150 \times 115 $ & 19 Nov 1998 & 17250  \\
   King 5 & V & $150 \times 085 $ & 22 Oct 1998 & 12750   \\
   Berkeley 20 & V & $150 \times 123 $ & 10 Dec 1998 & 18450  \\
               & V & $150 \times 130 $ & 15 Dec 1998 & 19500  \\
  \hline
  \end{tabular}
  \label{expo}
  \end{table*}


\section{Kinematical data and member selection}\label{kin}

Proper motion of any star provides velocity in two orthogonal
directions, which are used to determine the membership criteria of
the star in a star cluster (Sagar \& Bhatt 1989).
The second data
release of Gaia mission (Gaia DR2) provides five-parameter astrometric
data and photometric data in three filters. The astrometric data contains
celestial position, parallax and proper motion data of more than 1
billion sources (Gaia Collaboration {\em at al.} 2018).
We used Gaia proper motion and parallax data to select the cluster members and
photometric data in $G$, $G_{bp}$ and $G_{rp}$ filters for isochrone fitting.

\begin{figure}
\includegraphics[width=\columnwidth]{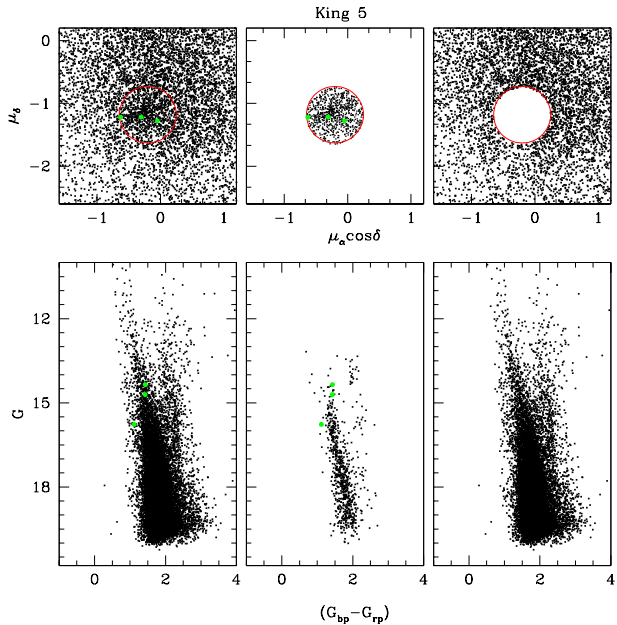}
\caption{Same as Fig. \ref{vpd_7}, but for the cluster King 5.}
\label{vpd_5}
\end{figure}

To separate cluster stars from field stars, we plotted the 
vector point diagrams (VPD) in $\mu_{\alpha}cos\delta$
and $\mu_\delta$ in top panels of the Fig. \ref{vpd_7} 
and \ref{vpd_5} for the clusters Berkeley 69, King 7 and King 5.
$G$ versus $(G_{bp}-G_{rp})$ colour-magnitude diagrams are plotted
in the lower panels. The VPDs of these clusters show a compact
distribution of stars as compared to the scattered stars.
We also plotted VPD for the cluster Berkeley 20 but could not find a separation
between field and cluster stars. This may be because of the cluster is at a 
large distance ($\sim$ 8 kpc).
The left panels of the figures represent 
total stars in our sample, middle panels show the stars with similar
proper motion and right panels show stars with different proper motion.
We drew a circle around the eye estimated centre of the stars in VPDs
to separate cluster members from field stars.
The chosen radius is a compromise between loosing cluster
members with poor proper motion and including field region stars.
The circle radii are taken as 0.4, 0.5 and 0.5 
mas $yr^{-1}$ for Berkeley 69, King 5 and King 7 respectively. 
We selected the stars in different magnitude bins based on the proper 
motion error described in Gaia Collaboration et al. (2018a). 
We calculated the mean parallax of the member stars for each cluster.
Mean parallax are found as $0.25\pm0.15$, $0.28\pm0.12$ and $0.34\pm0.13$
mas for the clusters Berkeley 69, King 7 and King 7 respectively.
Finally, we consider a star as member if it lie inside the circle in the VPD and
has parallax within 3$\sigma$ of the mean parallax of the cluster. The stars
shown in the middle
panels of the VPDs in Fig. \ref{vpd_7} and \ref{vpd_5} lie inside the
circle and have parallax within 3$\sigma$ of the mean parallax.
Selected members show a clear main sequence 
for all the clusters. Green points in these VPDs are the variable stars
detected in the cluster fields which is discussed in section \ref{var}
However in the VPD of Berkeley 69, cluster stars are not clearly distinguishable
from field stars and hence main-sequence includes field non members also.

\begin{figure}
\includegraphics[width=\columnwidth]{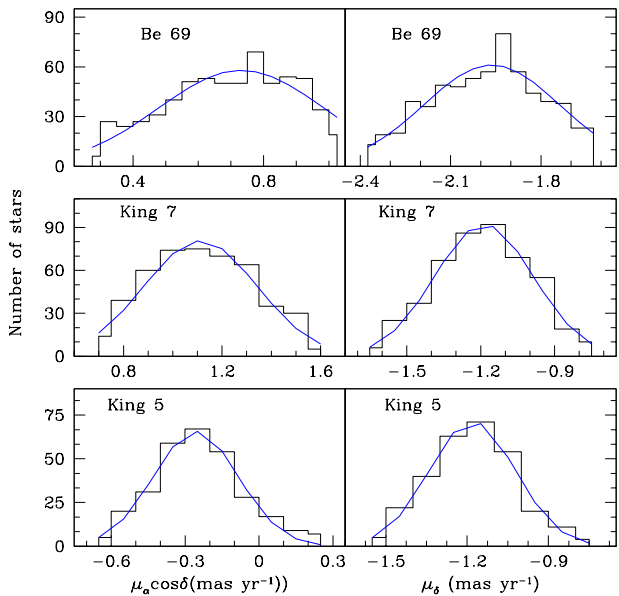}
\caption{Histograms in $\mu_{\alpha}cos\delta$ and $\mu_{\delta}$
for the clusters Berkeley 69, King 7 and King 5. The blue curve represents Gaussian 
fitting for the calculation of mean proper motion.}
\label{pradec}
\end{figure}

Mean proper motions in $\mu_{\alpha}cos\delta$ and $\mu_{\delta}$ of
the clusters Berkeley 69, King 7 and King 5 are calculated with the help of
histograms, shown in Fig.\ref{pradec}. For these histograms, we used
the most probable stars selected with the help of VPDs. Mean and 
standard deviation in $\mu_{\alpha}cos\delta$ and $\mu_{\delta}$ are 
calculated by fitting Gaussian function on the histograms.
From this fitting, the mean proper motions of the clusters are found as
$0.73 \pm 0.02$ and $-1.96 \pm 0.01$ mas for Berkeley 69, $1.11 \pm 0.02$ and 
$-1.19 \pm 0.01$ mas $yr^{-1}$ King 7 and
$-0.26 \pm 0.01$ and $-1.18 \pm 0.01$ mas $yr^{-1}$ for King 5 in 
RA and Dec directions respectively.


\section{Fundamental parameters of Berkeley 69, King 7 and King 5}\label{fpara}
\begin{figure*}
\includegraphics[width=6cm,height=6cm]{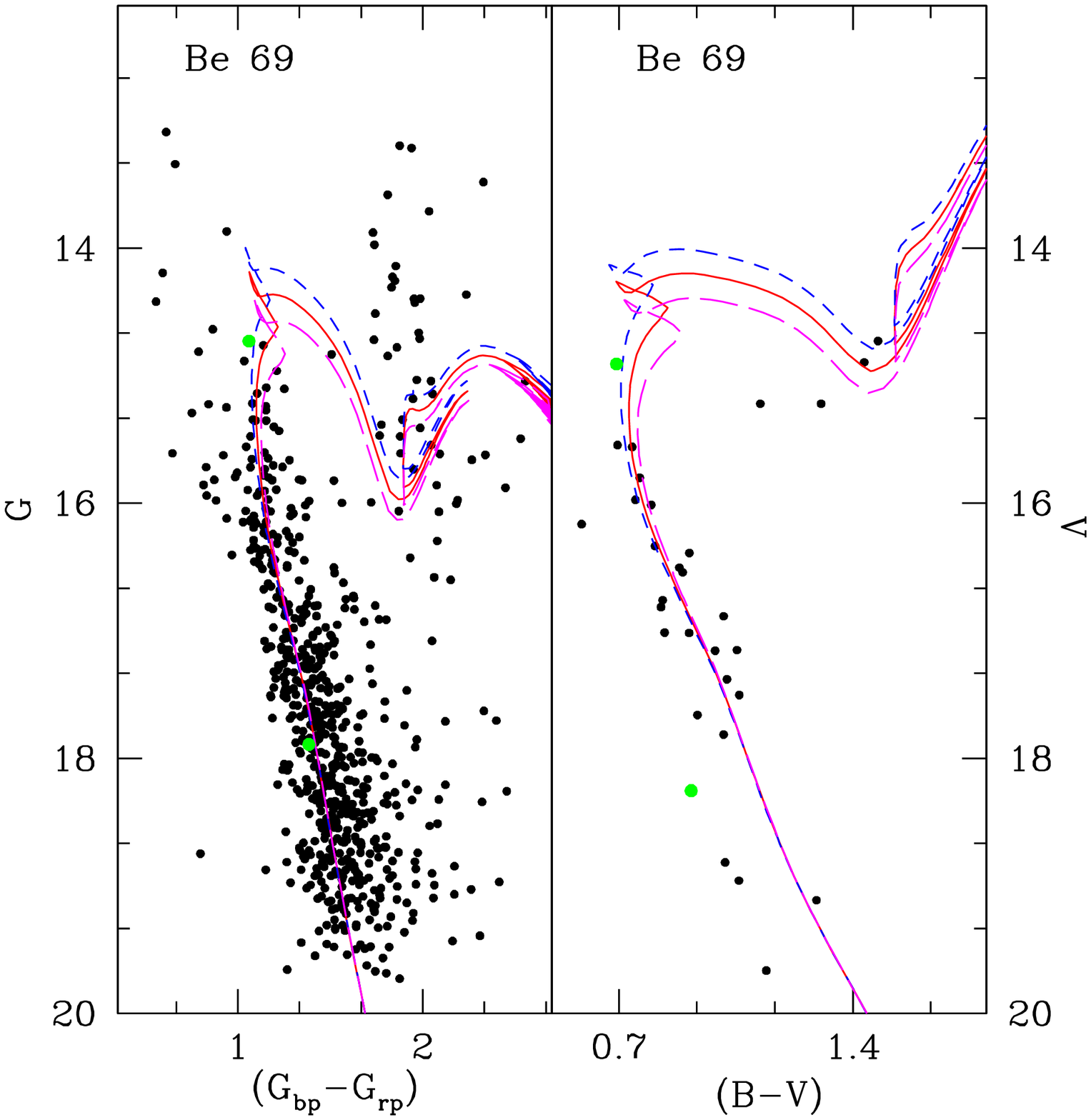}
\includegraphics[width=6cm,height=6cm]{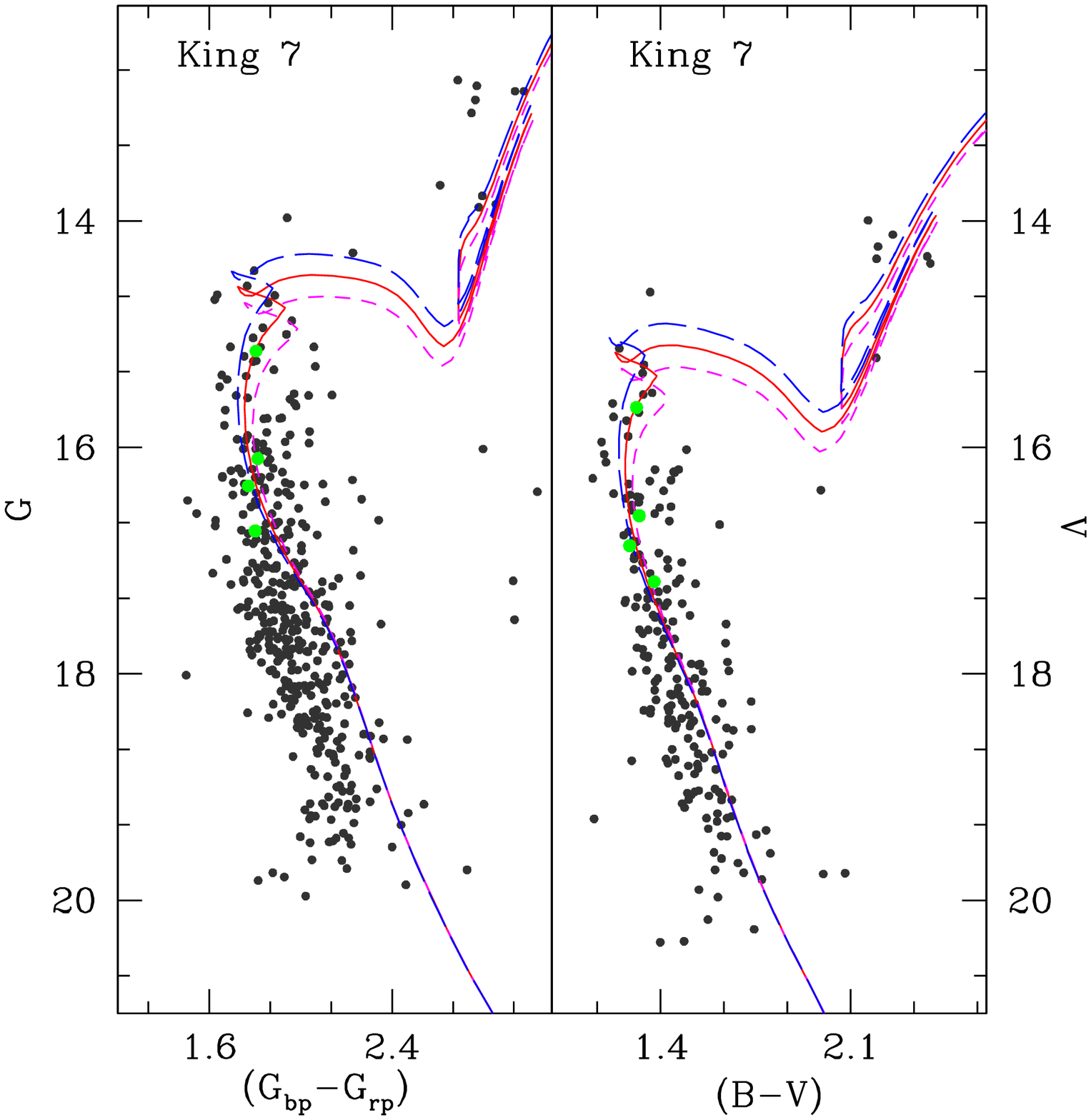}
\includegraphics[width=6cm,height=6cm]{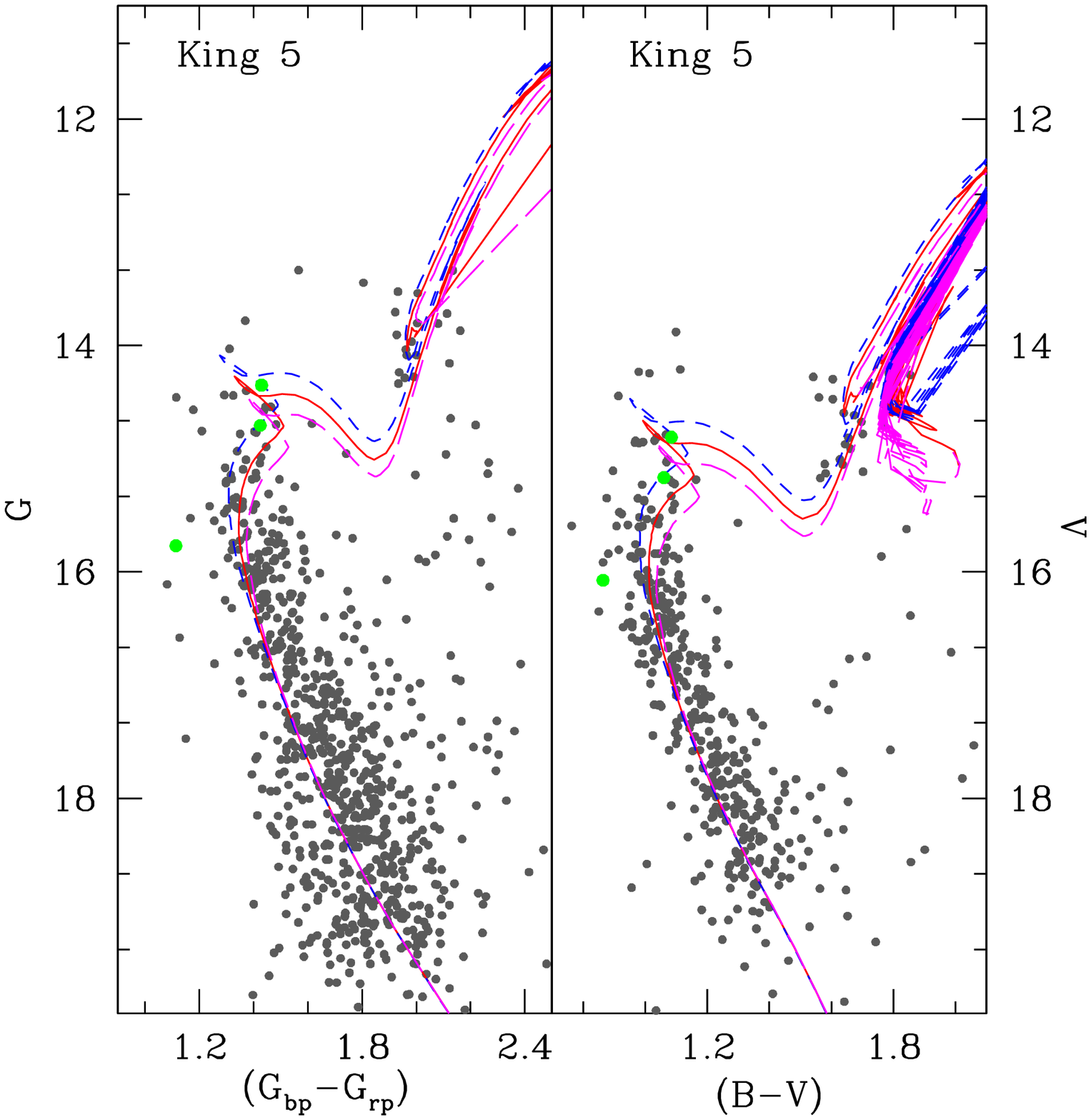}
\caption{Isochrones by Marigo {\em et al.} (2017) of $Z=0.008$ and age
$0.79 \pm 0.09$ Gyr, $Z=0.008$ and age $0.79 \pm 0.09$ Gyr, $Z=0.02$ and 
age $1.59 \pm 0.19$ Gyr 
Gyr are fitted on the ($G,G_{bp}-G_{rp})$ and ($V,B-V$) CMDs
for the clusters Berkeley 69, King 7 and King 5 respectively. The green dots denote the 
variable stars detected in the clusters and the red lines are the isochrones.}
\label{iso}
\end{figure*}

Fundamental parameters of the clusters Berkeley 69, King 7 and King 5 are reported in our previous studies (Durgapal {\em et al. 1996};
Durgapal {\em et al.} 1997; Durgapal {\em et al.} 1998; 
Durgapal {\em et al.} 2001).
In the present analysis, we revisited
the parameters using the most probable cluster members selected
with the help of Gaia DR2 proper motion data.
We have plotted $G$ versus ($G_{bp}-G_{rp}$) CMDs of Berkeley 69, King 7
and King 5. We have also plotted $V$ versus $(B-V)$
CMDs using common
stars between most probable cluster members selected by VPDs and
photometric data used in our previous papers for Berkeley 69, King 7 and King 5. 
The fitted isochrones are shown in Fig. \ref{iso}.
For isochrone fitting, we tried to fit several 
isochrones given by Marigo {\em et al.} (2017) on the cluster CMDs.
For Berkeley 69, we find that isochrones of $Z=0.008$ and age $0.79 \pm 0.09$ Gyr
show good fitting.
For King 7, isochrones of $Z=0.008$ and age 
$0.79 \pm 0.09$ Gyr are found to be satisfactorily fitted. For King 5, best fitted
isochrones are of $Z=0.02$ and age $1.59 \pm 0.19$ Gyr. The colour-excess
$E(B-V)$ and $E(G_{bp}-G_{rp})$ for Berkeley 69 are found as $0.68\pm0.03$ and $1.05\pm0.01$ mag.
For King 7 these are determined as $1.22\pm0.03$ and $1.70\pm0.02$
mag respectively. For King 5, these values are estimated as $0.63\pm0.02$ and $0.85\pm0.01$ mag
respectively. We also determined the distances of these clusters using 
the distance modulus in $G$ versus $(G_{bp}-G_{rp})$ plot as $2.53\pm0.02$,
$2.07\pm0.28$ and $2.08\pm0.02$ kpc for the clusters Berkeley 69, King 7 and 
King 5 respectively.
For interstellar extinction we adopted the value $A_{G}/E(G_{bp}-G_{rp})=
1.89\pm 0.02$ given by Wang \& Chen (2019). 
The values of age, colour-excess, distance and metallicity are in good 
agreement
with our previously determined values. The estimated parameters are also listed
in Table \ref{para}.

\begin{table*}
   \centering
   \caption{Fundamental astrophysical and kinematical parameters of Berkeley 69, King 7
    and King 5 determined in present analysis, where d is the
    heliocentric distance of the clusters.} 
   \begin{tabular}{ccccccccc}
   \hline\hline
  Cluster &  $\mu_{\alpha}cos\delta$ & $\mu_{\delta}$  & d & Age & $E(B-V)$ & Z \\
          &   mas $yr^{-1}$ & mas $yr^{-1}$ & kpc  & Gyr  & mag &    \\
  \hline
   Berkeley 69  & $0.73 \pm 0.02$  & $-1.96 \pm 0.01$ & $2.53 \pm 0.02$ & $0.79 \pm 0.09$ & $0.68\pm0.03$ & 0.008  \\
   King 7 &  $ 1.11 \pm 0.02 $ & $ -1.19 \pm 0.01 $ & $2.07 \pm 0.28$  &  $0.79 \pm 0.09$  & $1.22\pm0.03$ & 0.008  \\
King 5 &  $-0.26 \pm 0.01 $ & $ -1.18 \pm 0.01$ & $2.08 \pm 0.02$ & $1.59 \pm 0.19$ & $0.63\pm0.02$ & 0.02    \\
  \hline
  \end{tabular}
  \label{para}
  \end{table*}


\section{Variable stars in the direction of the clusters}\label{var}

Light curves of all the stars in the cluster region were examined to detect the variability.  A star was considered variable if its light curve shows a significantly (more than 3$\sigma$) larger dispersion than for the comparison stars.  A careful search indicates variability in a few stars in each cluster.
Locations of these detected variable stars are also shown in the cluster 
identification chart in Fig. \ref{id}, as well as in the CMD as shown in Fig. \ref{iso}. Since we did not have spectroscopic observations, we determined the
suspected spectral class of the variables with the help of standard data (Gray 2005).

For each variable star in these clusters, the differential photometric magnitude were obtained by subtracting comparison from the variable. Light curves of these variable stars are shown in the Figures \ref{fig:lc_be69}, \ref{fig:lc_king7}, \ref{fig:lc_king5} and \ref{fig:lc_be20}. The detailed information about these stars is given in Table \ref{tab:per}. A total of 4 stars in the cluster Berkeley 69, 7 stars in the cluster King 7, 5 stars in the cluster King 5 and 3 stars in the cluster Berkeley 20 were found to be variable.  In order to find the periodicity in these variables, we have performed a periodogram analysis by using the method of Lomb-Scargle periodogram (Lomb 1976; Scargle 1982).

\subsection{Berkeley 69} \label{b69}

\begin{figure*}
\begin{subfigure}{.33\textwidth}
\centering
\includegraphics[height=7cm,width=7cm,trim={0.2cm 0.0cm 0.0cm 0.0cm}]{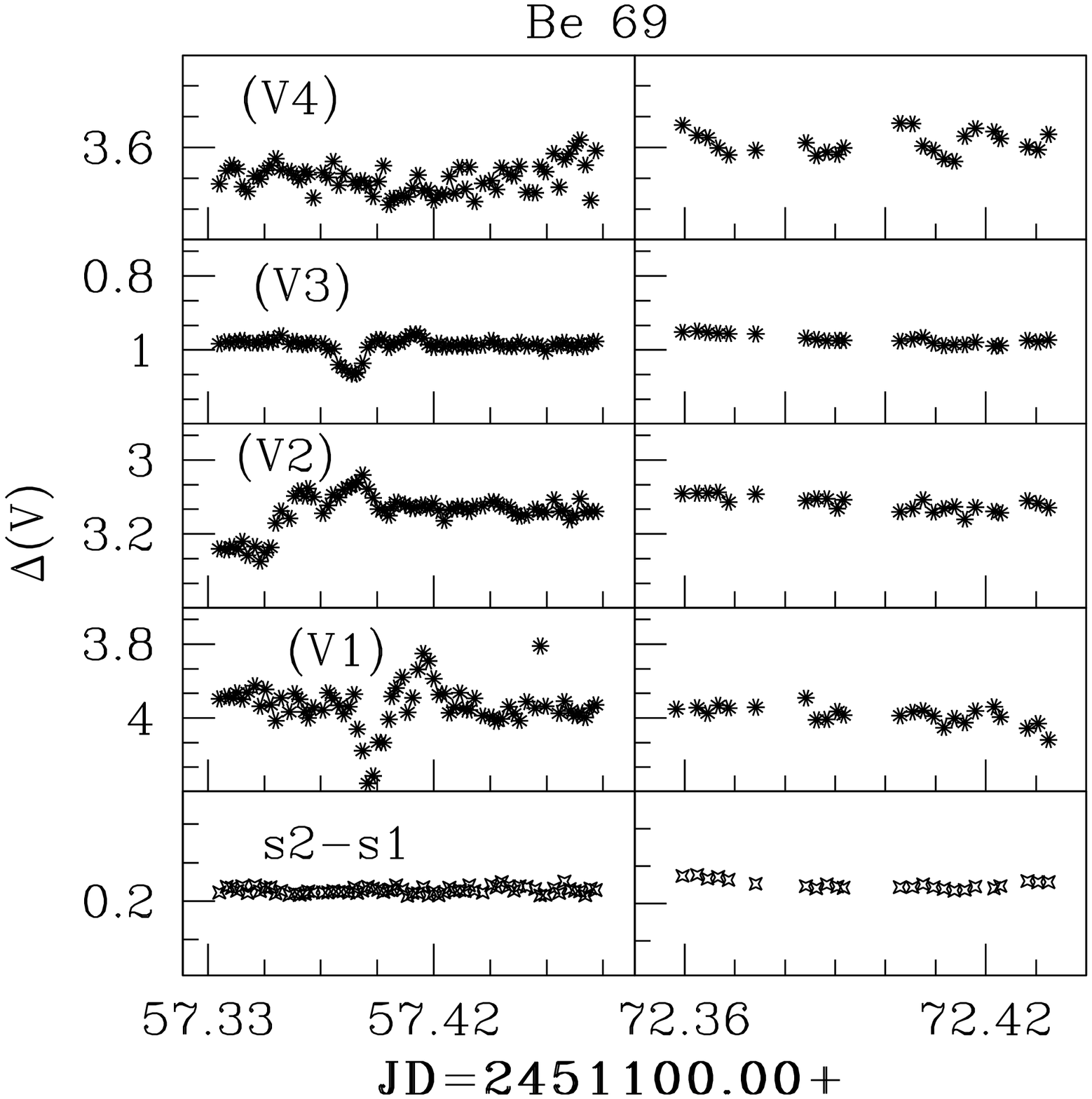}
\caption{Light curves}
\label{fig:lc_be69}
\end{subfigure}
\hspace{1.0cm}
\begin{subfigure}{.3\textwidth}
\centering
\includegraphics[height=5cm,width=5cm]{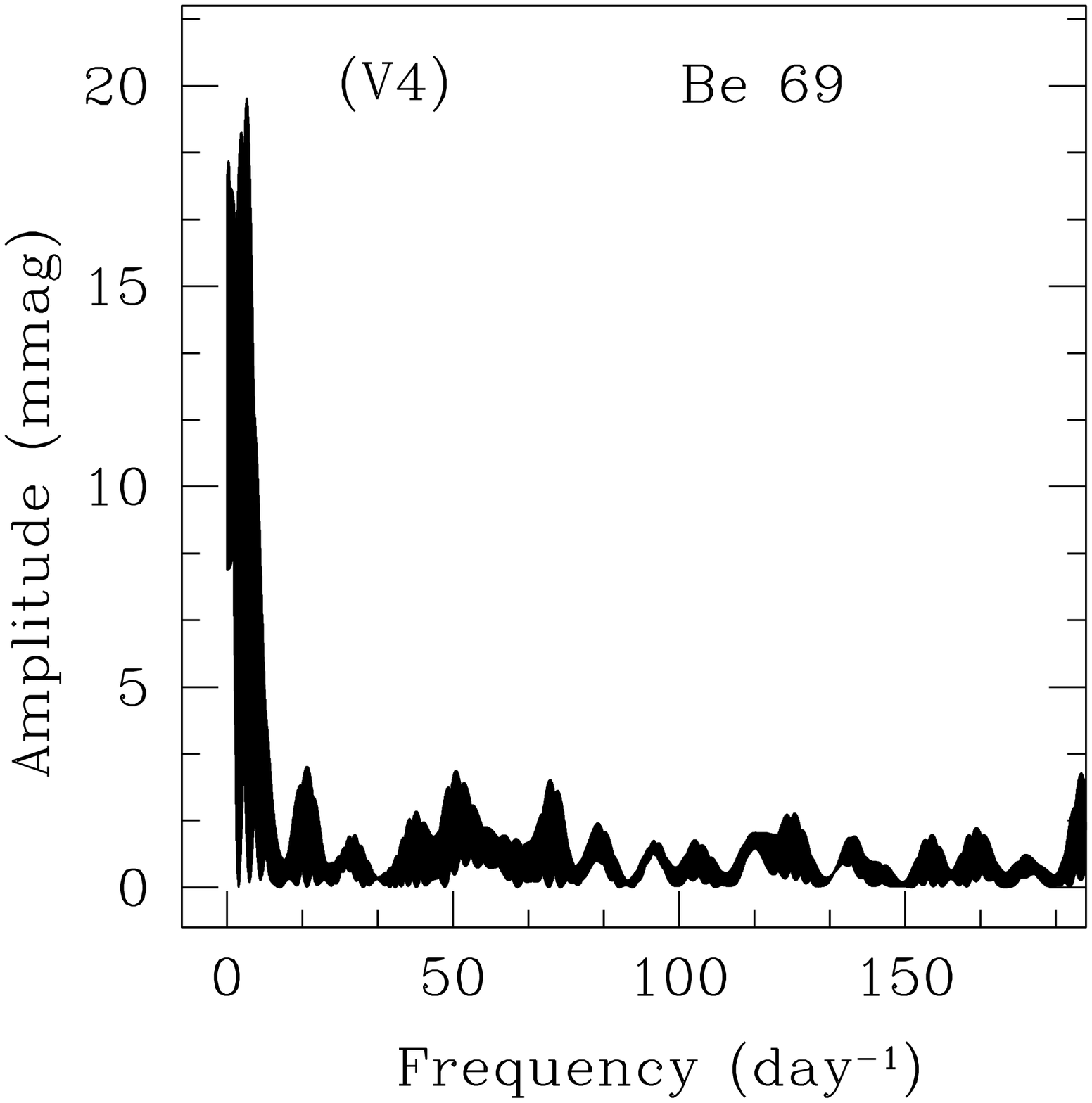}
\caption{Power spectra}
\label{fig:ps_be69}
\end{subfigure}
\begin{subfigure}{.3\textwidth}
\centering
\includegraphics[height=6cm,width=6cm]{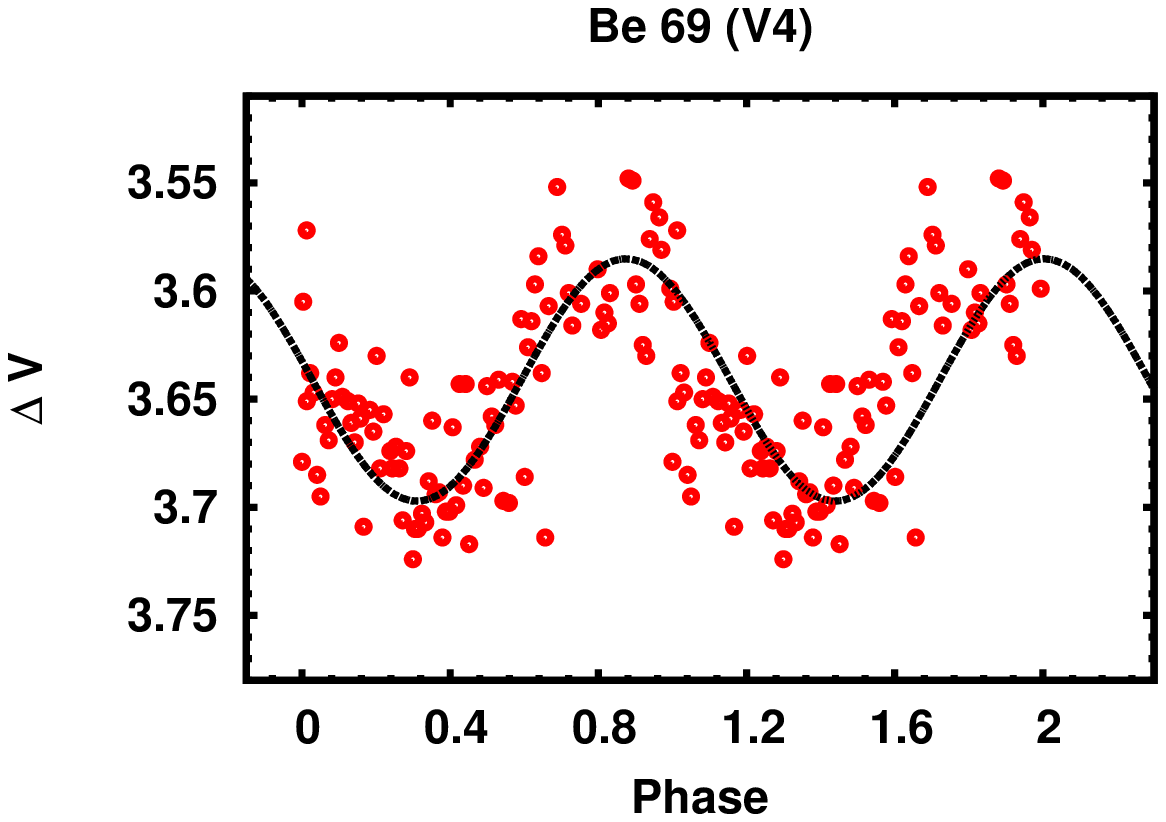}
\caption{ Folded light curve}
\label{fig:ph_be69}
\end{subfigure}
\caption{(a) Light curves of the variable stars in the cluster Berkeley 69.  Difference
of two comparisons (s2-s1) is also shown in the bottom panels. (b) Power spectra of the star V4 in the cluster Berkeley 69, and (c) Folded light curves of the star V4 along with the best fit sine curve.}
\label{fig:lfp_be69}
\end{figure*}

Berkeley 69 was observed in two nights on 9 December and 25  December 1998. 
A total of four stars were found to be variable in this cluster. The light 
curves of these variable stars along with differential light curve of 
comparison stars are shown in Fig. \ref{fig:lc_be69}. The average photometric 
error of the measurements as given by DAOPHOT is ~$\sim$ 0.02 mag.
All the stars which are detected as variables in the cluster Berkeley 69 are 
showing irregular behaviour in their light variations.  
No periodicity was seen in the light curves of V1, V2, and V3. However, 
the star V4 was found to be periodic variable. Figure \ref{fig:ps_be69} 
shows the power spectrum of the star V4, where the highest peak corresponds to 
a period of 0.225 d (or 5.4 hrs). In order to check whether this period is a real 
period, we have folded the data using an arbitrary ephemeris and period of 
0.225 d. Figure \ref{fig:ph_be69} shows the folded light curve of the stars 
V4 in the cluster Berkeley 69, where a clear signature of periodic variability was 
seen in this star. Further, a sine curve was also fitted in the folded light 
curve, which is shown by a curve in Figure \ref{fig:ph_be69}. The best 
fit sine curve also indicates the period of this star as 0.225 d. 
The location of the star V4 on the VPD of the cluster indicates
that this may be a field star with spectral type F0.  
Only V1 and V2 are cluster member stars because they have similar proper motion
as the cluster stars. Locations of these 
variables in the ID chart are also shown by the green circles in Figure \ref{id_69}.

\subsection{King 7}\label{k7}

\begin{figure*}
\begin{subfigure}{.33\textwidth}
\centering
\includegraphics[height=7cm,width=7cm]{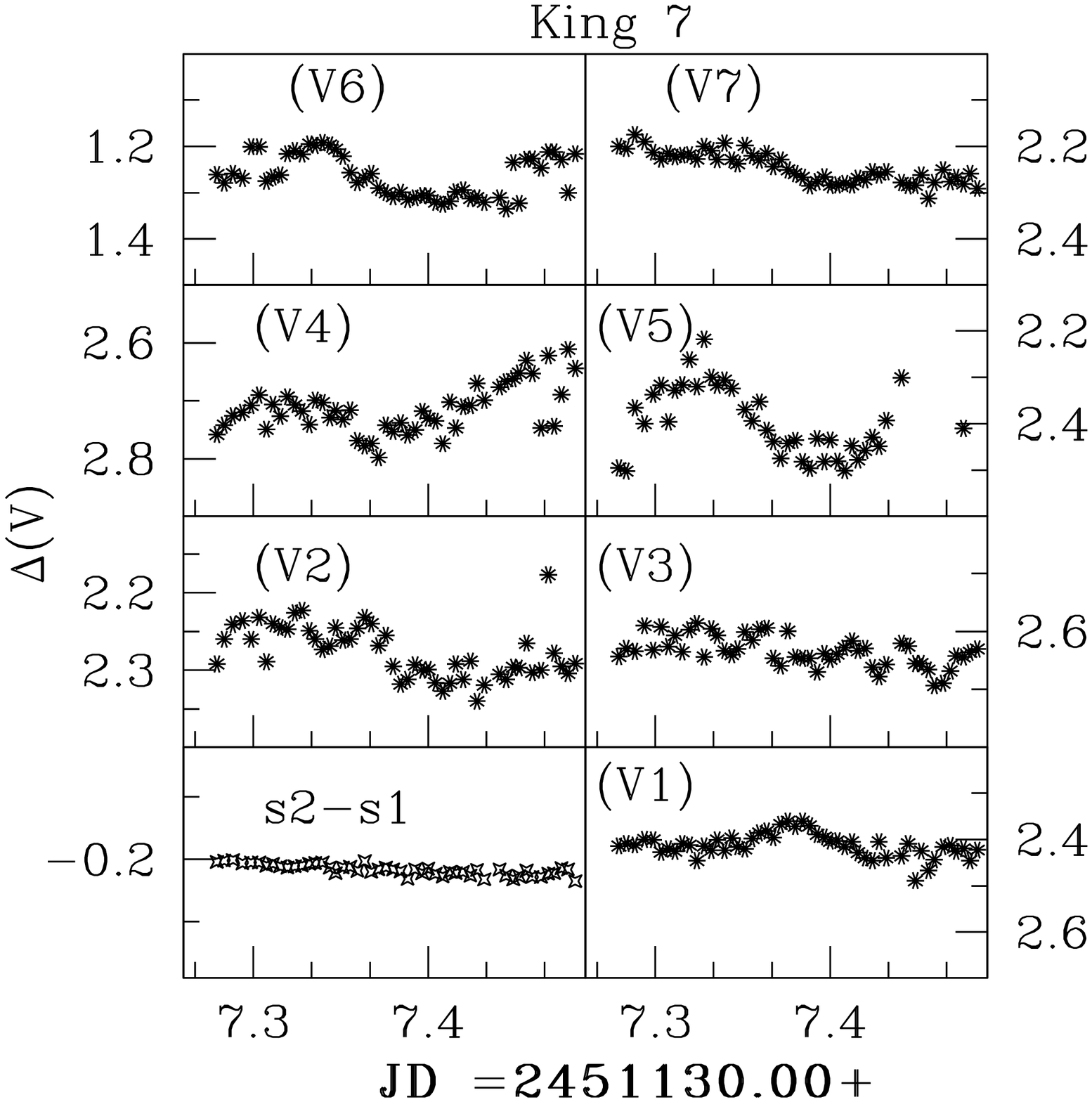}
\caption{Light curve}
\label{fig:lc_king7}
\end{subfigure}
\hspace{1.0cm}
\begin{subfigure}{.3\textwidth}
\centering
\includegraphics[height=5cm,width=5cm]{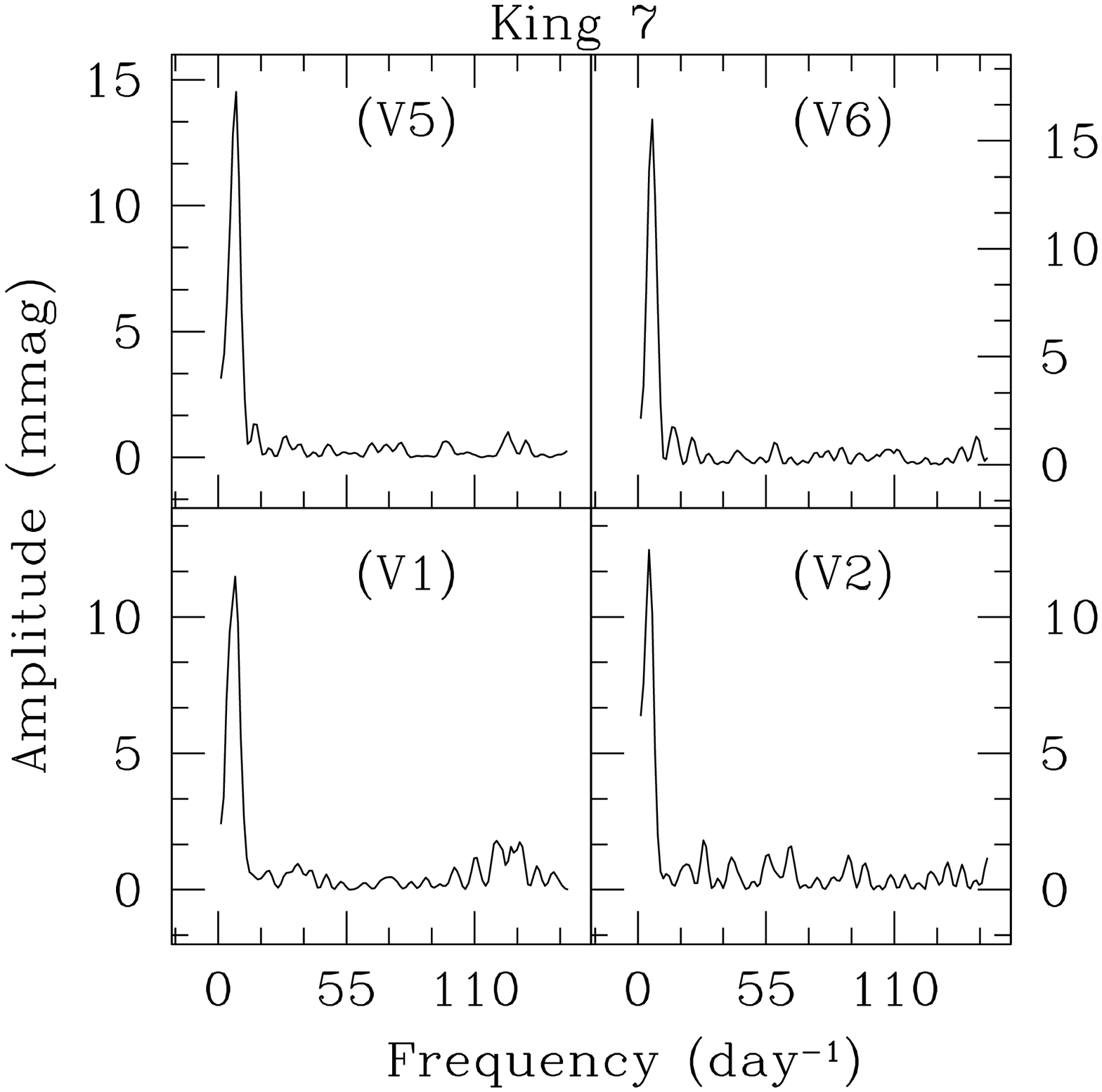}
\caption{Power spectra}
\label{fig:ps_king7}
\end{subfigure}
\begin{subfigure}{.3\textwidth}
\centering
\includegraphics[height=5.5cm,width=5.5cm]{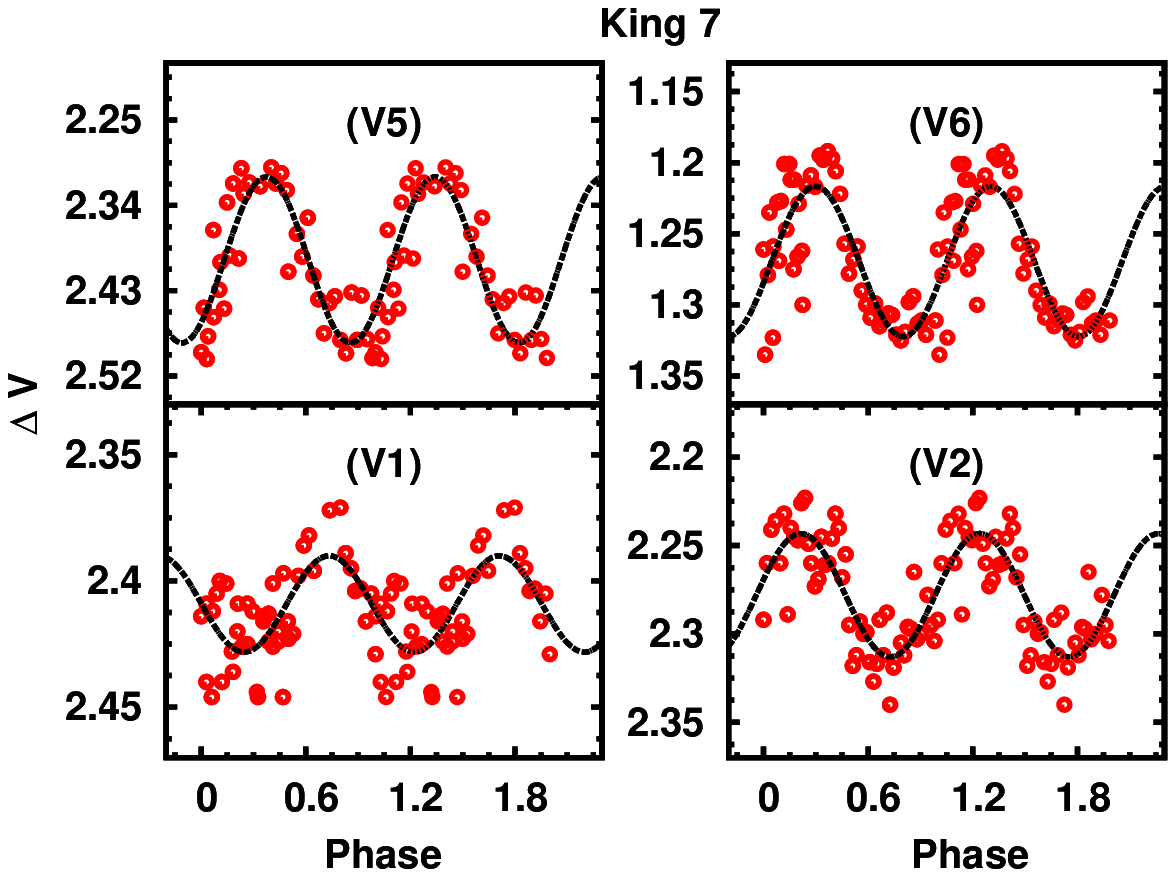}
\caption{Folded light curves}
\label{fig:ph_king7}
\end{subfigure}
\caption{(a) Light curves of the variable stars in the cluster King 7. Differential light curve of comparison stars is shown in the left bottom panel, (b) Power spectra of four stars V1, V2, V5, and V6 in the cluster King 7, and (c) Folded light curves of the stars V1, V2, V5,  and V6 along with the best fit sine curve.}
\label{fig:lfp_king7}
\end{figure*}

 Time series data of the cluster King 7 were taken in one night on 19 November 1998. The average photometric error was of the order of $\sim$0.015 mag. Using a similar approach as mentioned above, we detected a total of seven variable stars in this cluster (see also  Figure \ref{id}). The differential light curves of these seven variable stars are shown in Fig. \ref{fig:lc_king7} along with the differential light curve of comparison star. Periodogram analysis was carried out for all seven variable stars. Out of seven, only four stars  V1, V2, V5 and V6 were found to be periodic variable.  Figure \ref{fig:ps_king7} shows the power spectra of these 4 variables, where the highest peak in the power spectra show the period in the range of 0.13 to 0.21 d for these four stars (see Table \ref{tab:per}). To confirm the observed periodicity, we have folded the light curves of all four variables using an arbitrary ephemeris and the periods as mentioned in Table \ref{tab:per}. Figure \ref{fig:ph_king7} shows the folded light curves of these variables along with the best fit sine curve with the derived periods.   Spectral type of the all four periodic variable stars is estimated to be early A-type. The amplitude of variability ranges from 0.05 to 0.2 mag. Their periods, amplitude of variability and spectral type indicate that these are probable candidate of   $\delta$-Scuti type variables (see Breger 2000). Further, all these four stars were found to be member of the cluster King 7.  These periodic variables are also shown by green dots in the VPD and CMD  in Figure \ref{vpd_7}. Other three stars V3, V4 and V7 are found to be non-member and the periodicity in these stars could not be found due to lack of long term data.  Locations of all the seven variables in the ID chart are also shown in Figure \ref{id_7}.


\subsection{King 5} \label{k5}

\begin{figure*}
\begin{subfigure}{.33\textwidth}
\centering
\includegraphics[height=7cm,width=7cm]{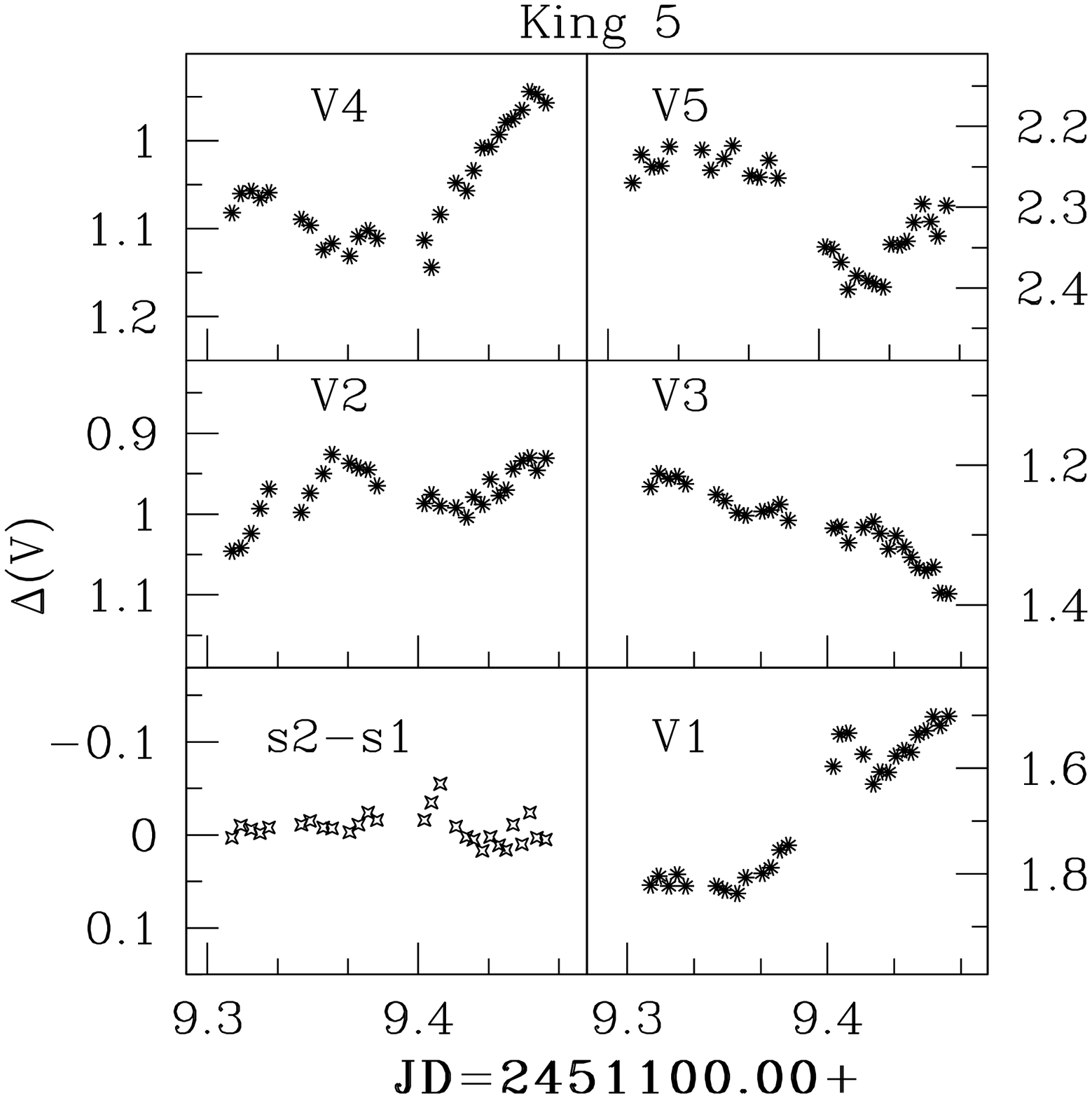}
\caption{Light curves}
\label{fig:lc_king5}
\end{subfigure}
\hspace{1.0cm}
\begin{subfigure}{.3\textwidth}
\centering
\includegraphics[height=5cm,width=5cm]{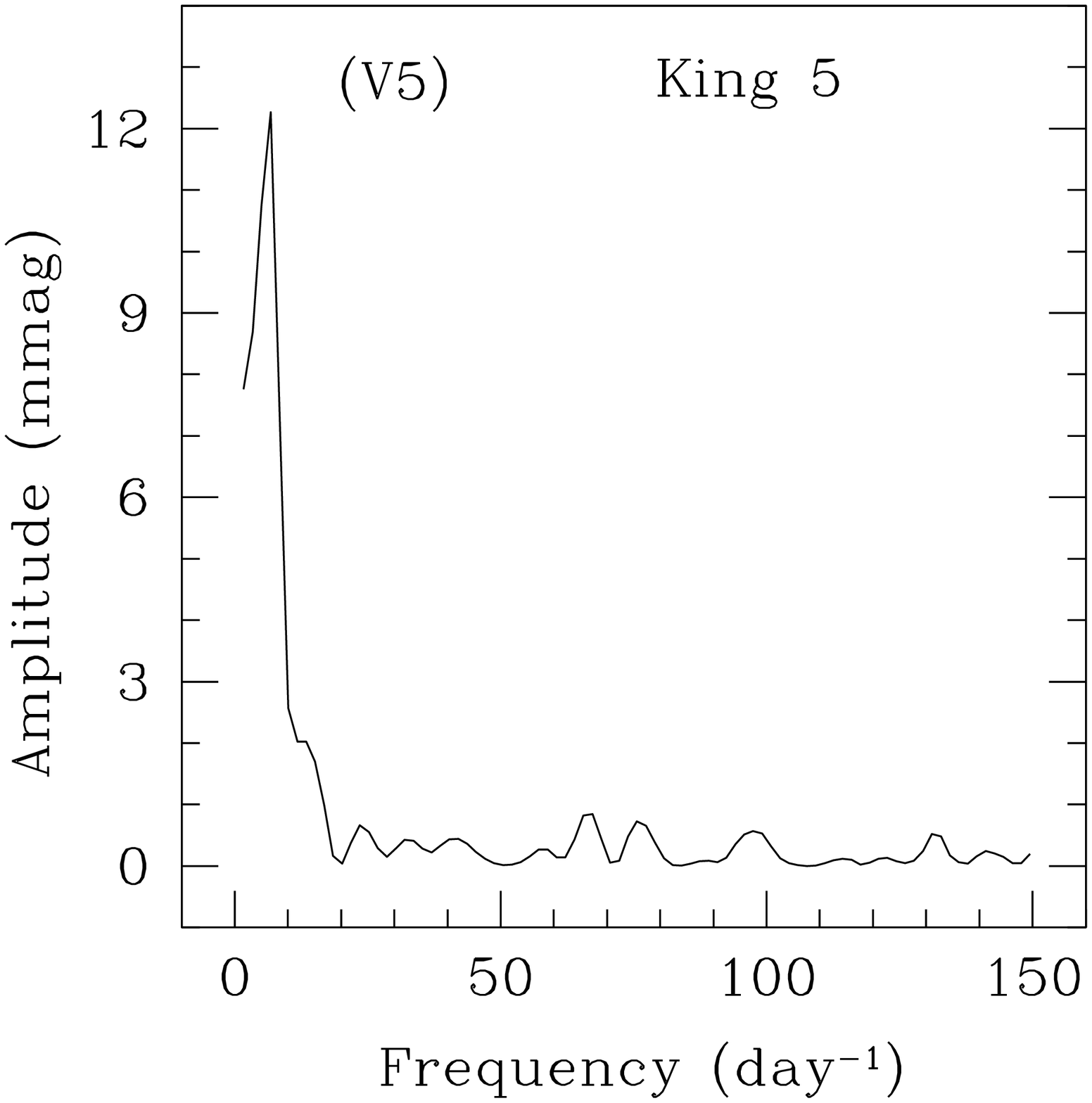}
\caption{Power spectra}
\label{fig:ps_king5}
\end{subfigure}
\begin{subfigure}{.3\textwidth}
\centering
\includegraphics[height=5.5cm,width=5.5cm]{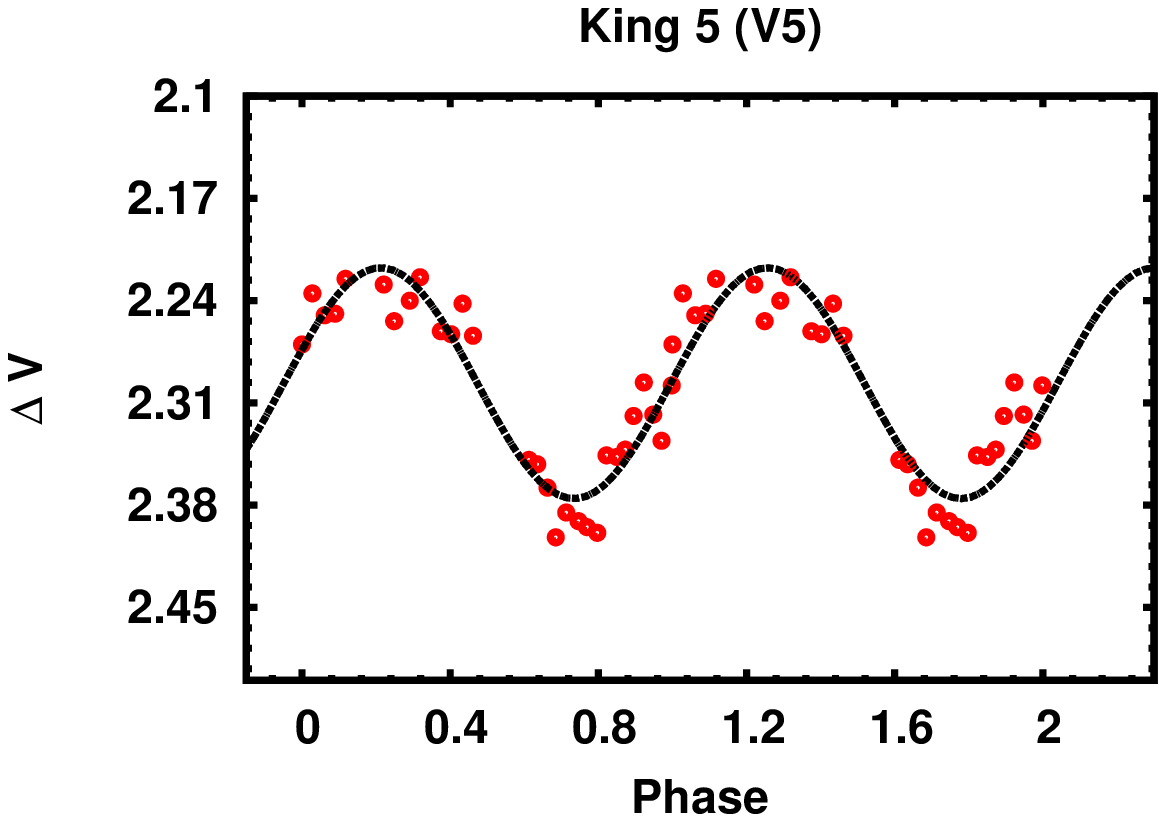}
\caption{Folded light curve}
\label{fig:ph_king5}
\end{subfigure}
\caption{(a) Light curves of the variable stars in the cluster King 5. Differential light curve of comparison stars is shown in the left bottom panel, (b) Power spectra of the star V5  in the cluster King 5, and (c) Folded light curves of the star V5 along with the best fit sine curve. }
\label{lfp_5}
\end{figure*}

The cluster King 5 was also observed for one night on 22 October 1998. The photometric error in determining
the magnitude of these stars was $\sim$ 0.015 mag. Our variability analysis revealed the variability in five stars of this cluster.  Light curves of these variable stars are shown in Figure \ref{fig:lc_king5}.  A periodogram analysis was performed to search the periodicity in these detected variables. We could not find variability period in any one of theses variable star stars. This could be due to their incomplete light curve, which is also evident from Figure \ref{fig:lc_king5}. However, the star V5 shows almost one cycle of observation, for which the highest peak in the power spectra showed a period of $\sim 0.15$ d (see Figure \ref{fig:ps_king5}). Folded light curve of the star V5 is shown in Figure \ref{fig:ph_king5}, where the continuous line curve shows the best fit sine function. The best fit sine curve also indicates that the period of the star V5 $\sim 0.15$ d.  The amplitude of variability  $\sim 0.15$ mag, period of 0.15 d ($\sim 3.6$ hr) and spectral type of A2 indicate that it is probably a $\delta$ Scuti type variable. 


\subsection{Berkeley 20} \label{b20}

\begin{figure*}
\begin{subfigure}{.33\textwidth}
\centering
\includegraphics[height=7cm,width=7cm]{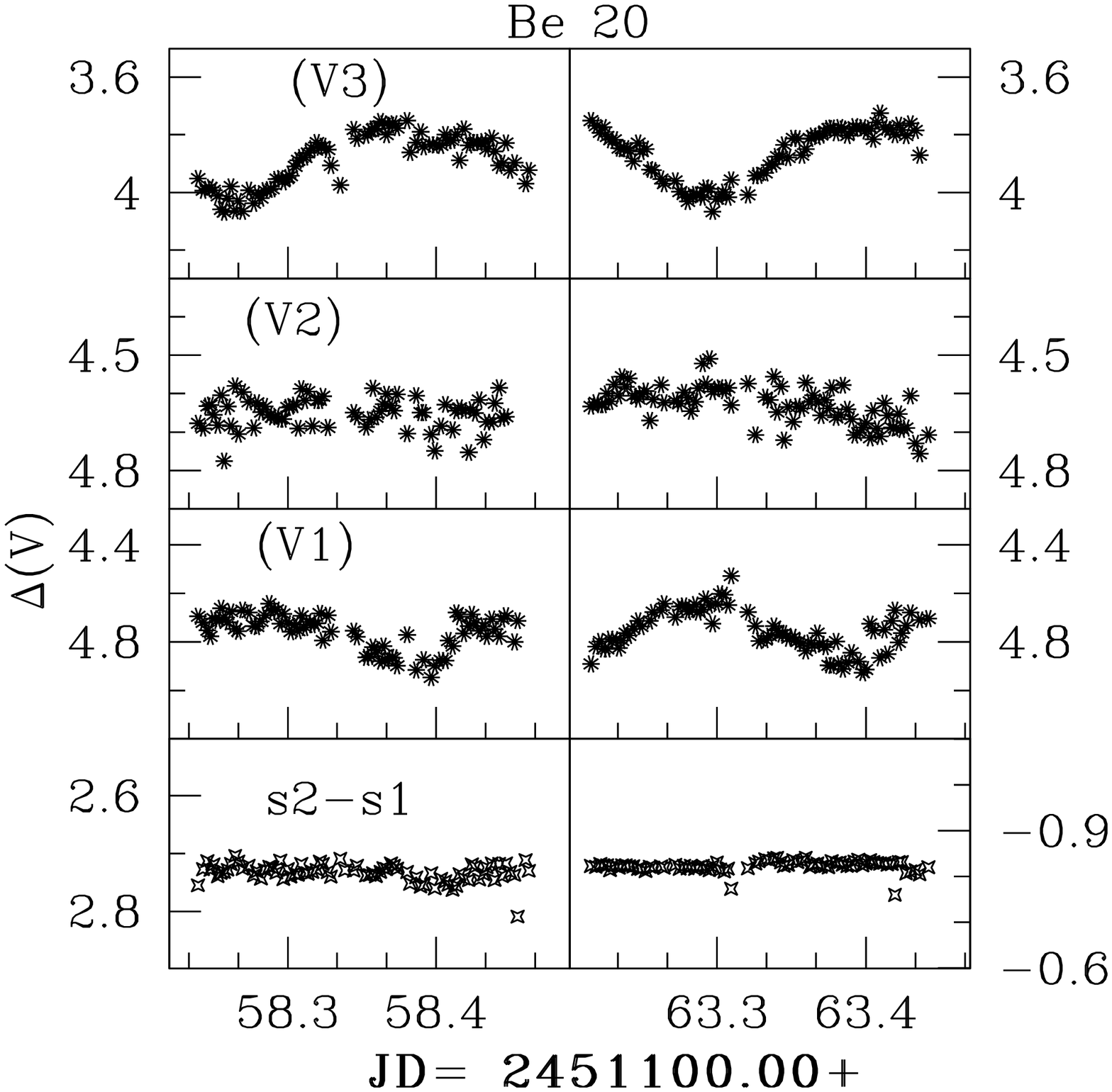}
\caption{Light Curves}
\label{fig:lc_be20}
\end{subfigure}
\hspace{1.0cm}
\begin{subfigure}{.3\textwidth}
\centering
\includegraphics[height=5cm,width=5cm]{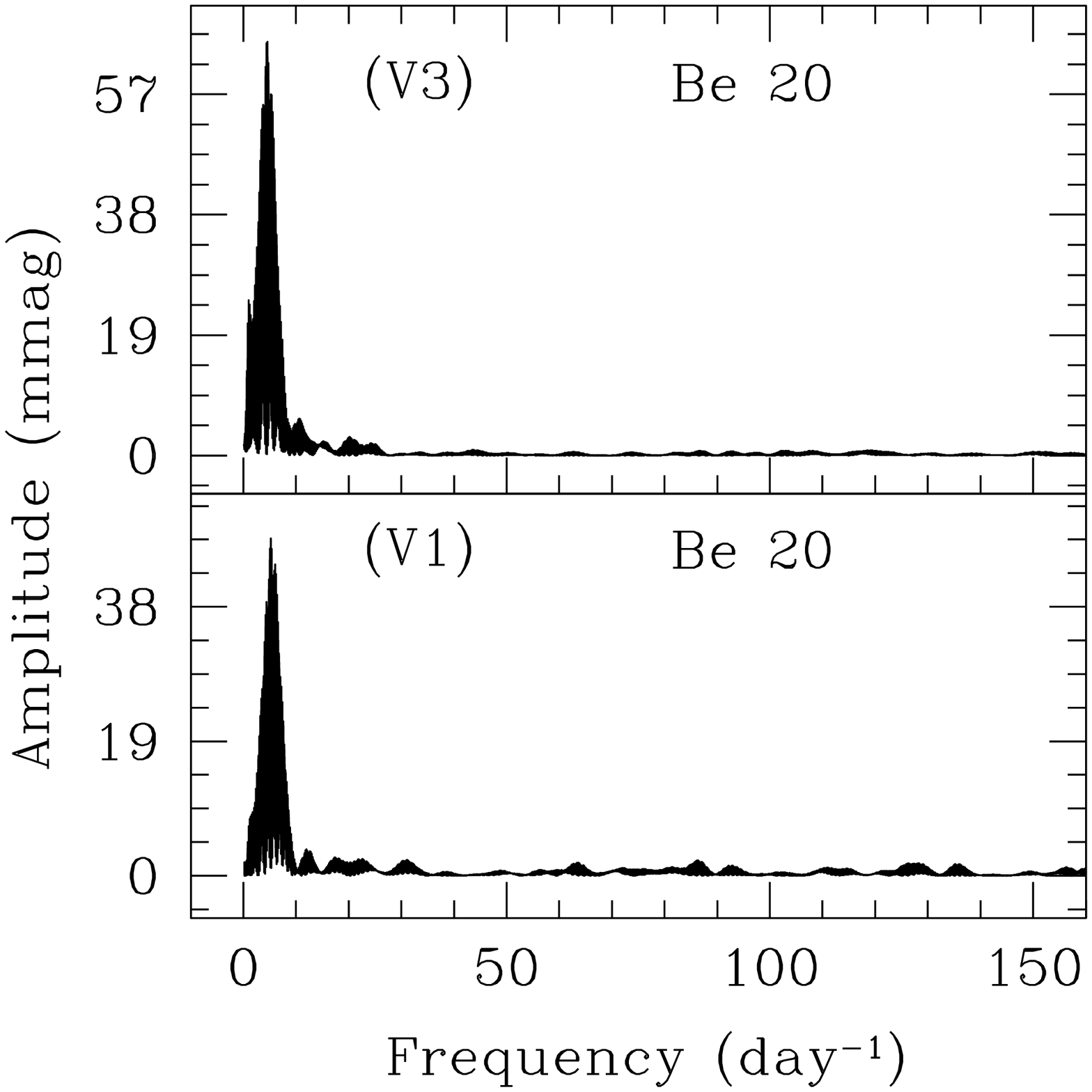}
\caption{Power spectra}
\label{fig:ps_be20}
\end{subfigure}
\begin{subfigure}{.3\textwidth}
\centering
\includegraphics[height=5.5cm,width=5.5cm]{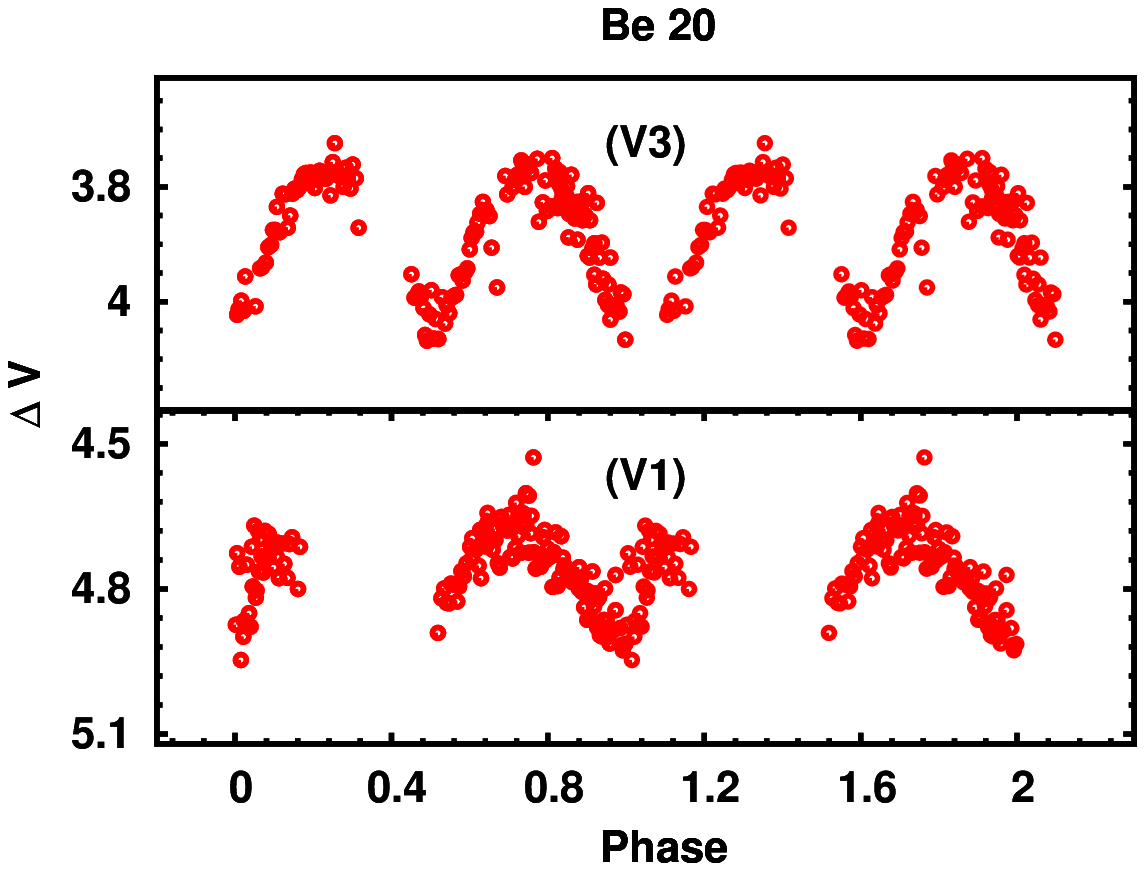}
\caption{Folded light curves}
\label{fig:ph_be20}
\end{subfigure}
\caption{(a) Light curves of the variable stars in the cluster Berkeley 20. Differential light curve of comparison stars is shown in the  bottom panels, (b) Power spectra of the stars V1 and V3  in the cluster Berkeley 20, and (c) Folded light curves of the stars V1 and V3.  }
\label{fig:lfp_be20}
\end{figure*}

\begin{table}
\tiny
\caption{ Details of variable stars in the clusters Berkeley 69, King 7, King 5 and Berkeley 20.  }
\label{tab:per}
\begin{tabular}{lcccccc}
\hline
Star&$\alpha$&$\delta$&M$_{v}$&(B-V)$_{0}$&Probable&Probable \\
&&&&&period&spect. \\
&&&&&(days)& class\\
\hline
\multicolumn{6}{c}{Berkeley 69}\\
V1* & 05:24:23.83 & +32:38:24.86 & 3.95 & 0.27 & -                &F0  \\
V2* & 05:24:25.17 & +32:38:16.34 & 3.15 & 0.34 & -                &F2  \\
V3 & 05:24:24.46 & +32:38:24.65 & 0.61 & 0.04 & -                &A1  \\
V4 & 05:24:20.64 & +32:37:07.17 & 3.96 & 0.24 & 0.225 $\pm$ 0.001&F0  \\
S2 & 05:24:28.04 & +32:37:13.43 & 0.43 & 0.82 & -                &-   \\
\multicolumn{6}{c}{ King 7}\\
V1*& 03:59:04.64 & +51:46:58.44 & 1.27 & 0.04 & 0.13 $\pm$ 0.02 & A2 \\
V2*& 03:59:06.75 & +51:45:47.93 & 1.00 & 0.07 & 0.21 $\pm$ 0.05 & A3 \\
V3 & 03:59:11.60 & +51:48:51.52 & 1.50 & 0.02 & -               & A1 \\
V4 & 03:59:28.07 & +51:49:40.42 & 0.79 & 0.71 & -               & G0 \\
V5*& 03:59:08.88 & +51:47:25.47 & 1.58 & 0.13 & 0.13 $\pm$ 0.02 & A5 \\
V6*& 03:59:04.35 & +51:46:17.00 & 0.05 & 0.06 & 0.16 $\pm$ 0.03 & A2 \\
V7 & 03:59:14.49 & +51:45:06.57 & 1.08 & 0.08 & -               & A3 \\
S2 & 03:59:09.51 & +51:45:36.54 &-1.26 & 0.94 & -               &    \\
\multicolumn{6}{c}{ King 5}\\
V1  & 03:14:54.15 & +52:42:08.27 & 1.57 & 0.33 &-                &-  \\
V2  & 03:14:39.87 & +52:44:04.74 & 1.02 & 0.27 &-                & A8\\
V3* & 03:14:30.55 & +52:41:35.53 & 1.17 & 0.24 & -               & A8\\
V4* & 03:14:42.38 & +52:40:02.41 & 0.84 & 0.19 & -               & A7\\
V5* & 03:14:27.76 & +52:40:50.81 & 2.07 & 0.04 & 0.15 $\pm$ 0.04 & A2\\
S2  & 03:14:46.00 & +52:41:21.80 &-0.18 & 0.92 &                 &   \\
\multicolumn{6}{c}{Berkeley 20}\\
V1 & 05:32:38.45 & +00:11:36.23 & 3.28 & 0.62 & 0.386 $\pm$ 0.002 & G0\\
V2 & 05:32:36.80 & +00:14:13.76 & 3.31 & 0.75 &  -                & G2\\
V3 & 05:32:35.82 & +00:11:19.74 & 2.26 & 0.58 & 0.438 $\pm$ 0.002 & G0\\
S2 & 05:32:33.63 & +00:12:33.39 & 0.63 & 0.97 &        -          & - \\
\hline
\end{tabular}
~~* Cluster member.
\end{table}
Time series observations of this cluster were taken in two nights with an
interval of five days on 10 December 1998 and  15 December 1998. Average photometric error as given by DAOPHOT was $\sim$ 0.012 mag. Using a similar approach as mentioned above in section 5.1, a total of three variable stars were identified in this cluster. Locations of these variable stars are also marked in Figure \ref{id_20}, whereas Figure \ref{fig:lc_be20} shows their light curve along with the light curve of the comparison star. 
Periodogram analysis was carried out for all the variable stars detected in the cluster Berkeley 20. No significant peak was found for the star V2 in the power spectra, whereas for the star V1 and V3 significant peaks at periods of 0.193 and 0.219 days  were found in their respective power spectra as shown in Figure \ref{fig:ps_be20}. Spectral types of these two variable stars were estimated as of G-type indicating these variables to be  W UMa-type binary. Power spectra of W UMa type of binaries show the peak power at the half of its original frequency due to the fact that they consist of two minima in one cycle of their light curves. Therefore, the period of the stars V1 and V3 were estimated as 0.386
and 0.438 days, respectively. Light curve of the star V1 was folded using the ephemeris HJD = 2451158.391+0.396 E, whereas for the stars for the star V3, we have used the ephemeris JD = 2451163.298 + 0.438 E. The JD corresponding to a phase zero was determined by fitting the second order polynomial to the minimum observed. Figure \ref{fig:ph_be20} shows the folded light curves of both the stars. The amplitude (minimum-maximum) of variation was found to be $\sim$ 0.3 mag for each V1 and V3. 
The current analysis suggests that the stars V1 and V3 in the cluster Berkeley 20 show the variability of W UMa type binaries.

\begin{figure}
\includegraphics[width=\columnwidth]{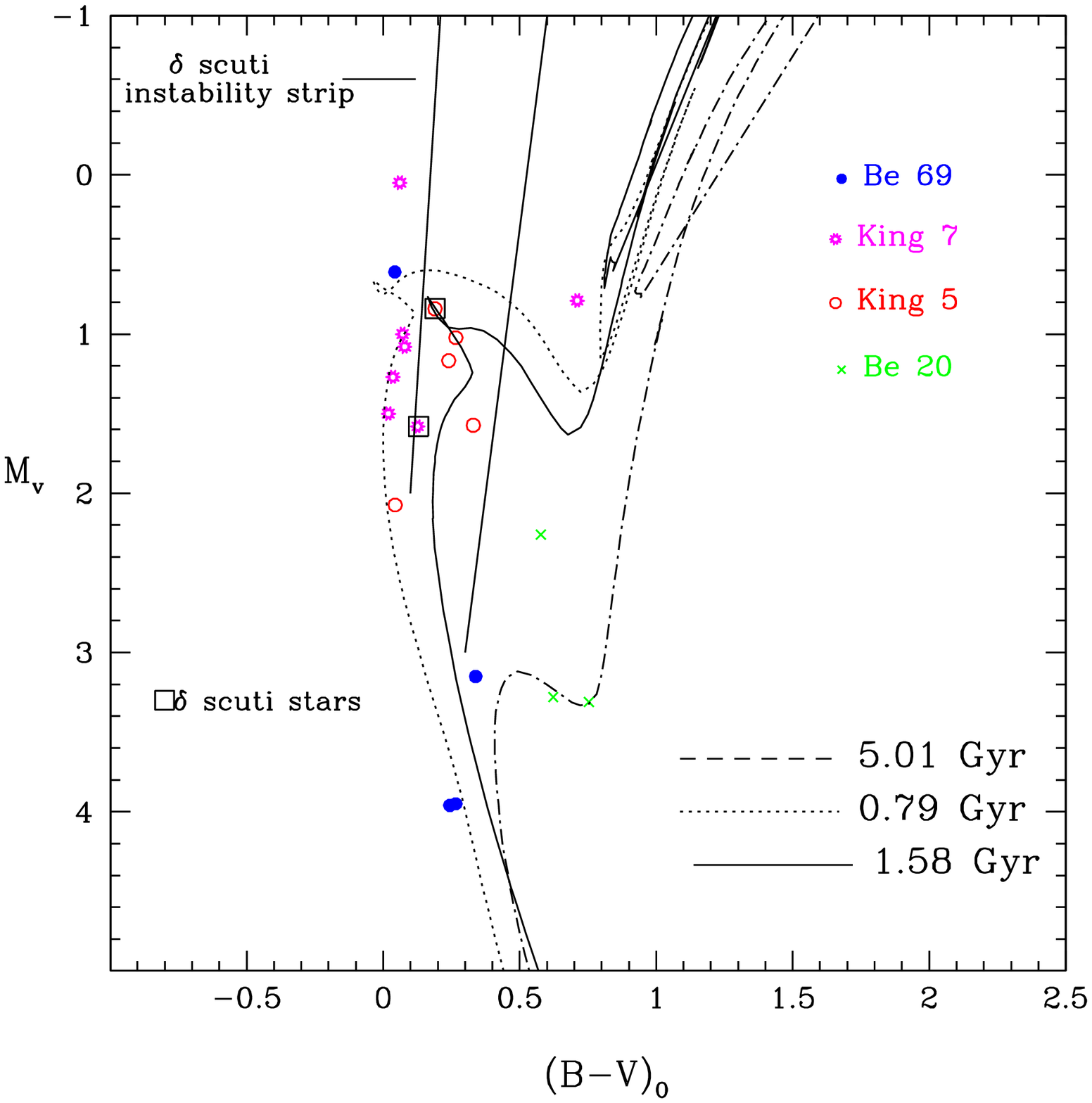}
\caption{The composite $M_{v}$, $(B-V)_{0}$ CMD of the clusters under discussion.
The position of detected variable stars and $\delta$ Scuti instability strip
are also shown in the figure.}
\label{cmd}
\end{figure}

Fig. \ref{cmd} shows the composite $M_{V}$, $(B-V)_{0}$ CMD of the clusters 
with detected variable stars.
$\delta$ Scuti instability strip and isochrones of different ages given by
Marigo et al. (2017) are also plotted in the figure. Two out of four
variables detected in King 5 within instability strip show light curves comparable to the
$\delta$ Scuti star. On the other hand one star of Be 69 and one star of King 5 and
five stars of King 7 are lying near the $\delta$ Scuti instability strip.
Based on the position of the variable stars in the CMD, their suspected
spectral class and their period, we can say that a few of them may be 
$\delta$ Scuti stars.

\section{Mass segregation}\label{mass}

\begin{figure}
\includegraphics[width=\columnwidth]{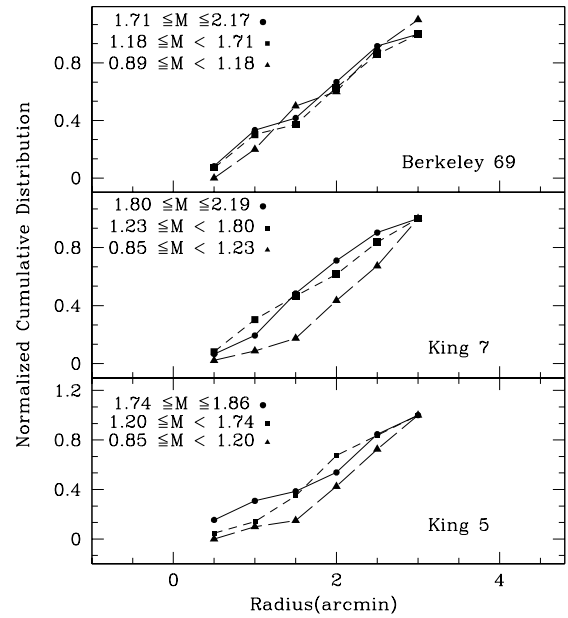}
\caption{Cumulative radial distribution of stars in the clusters
Berkeley 69, King 7 and King 5.
}
\label{mseg}
\end{figure}

Various studies in the literature have shown that massive stars are located 
towards the center of the cluster as compared to the less massive
stars. This configuration may be due to the star formation processes or due to the dynamical 
evolution in the cluster. We study the mass segregation by assuming the cluster 
members selected in this study. We divided the stars in different mass range and 
plotted the cumulative distribution as shown in Fig. \ref{mseg}.
This figure shows that mass segregation is present in the clusters
king 7 and King 5. We also performed the Kolmogorov-Smirnov (KS) test
to check whether these curves belong to the same population or not.
We found that in the case of King 7 and king 5, the curves belong to the different
population with a confidence level of 99 and 95 \% respectively whereas
for Berkeley 69 the confidence level is just 30 \%. This shows that King 7 and King 5 
have influenced by mass segregation.

\section{Conclusions} \label{con}

We have carried out time series photometry of the four clusters Berkeley 69, 
King 7, King 5, and Berkeley 20. A total of 19 variable stars (4 in Berkeley 69, 7 in 
King 7, 5 in King 5, and 3 in Berkeley 20) are identified in these clusters. 
Periodogram analysis of these variable stars showed the periodicity in eight 
of them with the period range of 0.13 - 0.43 days. In the other eleven stars, 
we could not find the periodicity due to the lack of sufficient data. Four 
stars in the cluster King 7 and one star in the cluster King 5 show the 
$\delta$ Scuti type of variability, whereas two stars in the cluster Berkeley 20 
have shown W UMa type of the variability. 
We believe that this preliminary identification of new variable stars 
in these four clusters will help for making a case for further observations 
that could inform more detailed properties of these variables. Mass 
segregation is observed in King 7 and King 5 clusters.

\section*{Acknowledgments}
This work has been partially supported by the Natural Science Foundation of China (NSFC-11590782, NSFC-11421303). This work has made use of data from the European Space Agency (ESA) mission GAIA processed by Gaia Data processing  and Analysis Consortium (DPAC), (https://www.cosmos.esa.int/web/gaia/dpac/consortium). 

\vspace{-1em}




\begin{theunbibliography}{} 
\vspace{-1.5em}



\bibitem{latexcompanion}
Breger, M., ASP Conference Series, Vol. 210, page 3. 

\bibitem{latexcompanion}
Bukowiecki, L., Maciejewski, G. 2008, Information Bulletin on Variable Stars, 5857, 1

\bibitem{latexcompanion}
Durgapal, A.K. 2001, Bulletin of the Astronomical Society of India, 29, 389

\bibitem{latexcompanion}
Durgapal, A.K., Mohan, V., Pandey, A.K., Mahra, H.S. 1996, 
Bulletin of the Astronomical Society of India, 24, 701

\bibitem{latexcompanion}
Durgapal, A.K., Pandey, A.K. 2001, A\&A, 375, 840

\bibitem{latexcompanion}
Durgapal, A.K., Pandey, A.K., Mohan, V. 1997, 
Bulletin of the Astronomical Society of India, 25, 489

\bibitem{latexcompanion}
Durgapal, A.K., Pandey, A.K., Mohan, V., 1998, 
Bulletin of the Astronomical Society of India, 26, 551

\bibitem{latexcompanion}
Durgapal, A.K., Pandey, A.K., Mohan, V. 2001, A\&A, 372, 71

\bibitem{latexcompanion}
Dutta, Somnath, Mondal, Soumen, Joshi, Santosh, Das, Ramkrishn, 2019, MNRAS, 487, 1765

\bibitem{latexcompanion}
Gaia Collaboration et al. 2018a, A\&A, 616, A1

\bibitem{latexcompanion}
Gaia Collaboration, Katz, D., Antoja, T., Romero-G´omez, M., Drimmel, R., Reyl´e,
C., Seabroke, G.M., Soubiran, C., Babusiaux, C., Di Matteo, P.,{\em et al.}  2018,
A\&A, 616, A11

\bibitem{latexcompanion}
Gray, D.F., 2005, The Observation and Analysis of Stellar Photospheres, 
3rd edition, Cambridge University Press, Cambridge, p 52

\bibitem{latexcompanion}
Hoq, S., Clemens, D.P. 2015, AJ, 150, 135

\bibitem{latexcompanion}
Hutchens, Z.L., Barlow, B.N., Soto, A.V., Reichart, D.E., Haislip, J.B., Kouprianov,
V.V., Linder, T.R., Moore, J.P. 2017, Open Astronomy, 26, 252 

\bibitem{latexcompanion}
Jeon, Y.B., Park, Y.H., Lee, S.M. 2016, 
Publication of Korean Astronomical Society, 31, 43


\bibitem{latexcompanion}
Joshi, Y. C., Joshi, S., Kumar, Brijesh, Mondal, Soumen, Balona, L. A., 2014, MNRAS, 437, 804

\bibitem{latexcompanion}
Lata, Sneh, Yadav, Ram Kesh, Pandey, A. K., Richichi, Andrea, Eswaraiah, C., Kumar, Brajesh, Kappelmann, Norbert, Sharma, Saurabh, 2014, MNRAS, 442, 273

\bibitem{latexcompanion}
Lata, Sneh, Pandey, A. K., Panwar, Neelam, Chen, W. P., Samal, M. R., Pandey, J. C.,  2016, MNRAS, 456, 2505

\bibitem{latexcompanion}
Lomb, N. R. 1976, Ap\&SS, 39, 447

\bibitem{latexcompanion}
Marigo, P., Girardi, L., Bressan, A., Rosenfield, P., Aringer, B., Chen, Y., Dussin,
M., Nanni, A., Pastorelli, G., Rodrigues, T.S., Trabucchi, M., Bladh, S., Dalcanton,
J., Groenewegen, M.A.T., Montalb´an, J., Wood, P.R. 2017, ApJ, 835, 77

\bibitem{latexcompanion}
Oralhan, ˙I.A., Karatas, Y., Schuster, W.J., Michel, R., Chavarr´ıa, C. 2015,
New A, 34, 195

\bibitem{latexcompanion}
Pandey, A.K., Durgapal, A.K., Bhatt, B.C., Mohan, V., Mahra, H.S. 1997, A\&AS, 122, 111

\bibitem{latexcompanion}
Popov, A.A., Zubareva, A.M., Burdanov, A.Y., Krushinsky, V.V., Avvakumova, E.A.,
Ivanov, K. 2017, Peremennye Zvezdy Prilozhenie, 17, 3

\bibitem{latexcompanion}
Riess, A.G., Casertano, S., Yuan, W., Macri, L., Bucciarelli, B., Lattanzi, M.G.,
MacKenty, J.W., Bowers, J.B., Zheng, W., Filippenko, A.V., Huang, C., Anderson,
R.I. 2018, ApJ, 861, 126
and Application to Gaia DR2: Implications for the Hubble Constant. ApJ 861, 126

\bibitem{latexcompanion}
Sagar, R., Bhatt, H.C. 1989, MNRAS, 236, 865


\bibitem{latexcompanion}
Scargle, J. D. 1982, ApJ, 263, 835

\bibitem{latexcompanion}
Schaefer, B.E., Bentley, R.O., Boyajian, T.S., Coker, P.H., Dvorak, S., Dubois, F.,
Erdelyi, E., Ellis, T., Graham, K., Harris, B.G., Hall, J.E., James, R., Johnston, S.J.,
Kennedy, G., Logie, L., Nugent, K.M., Oksanen, A., Ott, J.J., Rau, S., Vanaverbeke,
S., van Lieshout, R., Wyatt, M. 2018, MNRAS, 481, 2235

\bibitem{latexcompanion}
Smith, N., Aghakhanloo, M., Murphy, J.W., Drout, M.R., Stassun, K.G., Groh, J.H.
2019, MNRAS, 488, 1760



\bibitem{latexcompanion}
Stetson, P.B. 1987, PASP, 99, 191

\bibitem{latexcompanion}
Wang, S. \& Chen, X. 2019, ApJ, 877, 116

\bibitem{latexcompanion}
Yepez, M.A., Arellano Ferro, A., Schr ¨oder, K.P., Muneer, S., Giridhar, S., 
Allen, C. 2019, New A, 71, 1


\end{theunbibliography}
\end{document}